\documentclass[12pt]{article}

\begin{document}

\title{Biconformal supergravity and the AdS/CFT conjecture}
\author{{Lara B. Anderson\thanks{%
Department of Physics, Utah State University, Logan, Utah 84322} \thanks{%
Electronic mail: lbanderson@cc.usu.edu}} and {James T. Wheeler \thanks{%
Department of Physics, Utah State University, Logan, Utah 84322} \thanks{%
Electronic mail: jwheeler@cc.usu.edu}}}
\maketitle

\begin{abstract}
Biconformal supergravity models provide a new gauging of the superconformal
group relevant to the Maldacena conjecture. Using the group quotient method
to biconformally gauge $SU(2,2|N)$, we generate a 16-dim
superspace. We write the most general even- and odd-parity actions linear in
the curvatures, the bosonic sector of which is known to descend to general
relativity on a 4-dim manifold.
\end{abstract}

\section{\protect\smallskip Introduction}

For the past half decade, studies of $M/$string theory have been dominated
by interest in the relationships between string on specified backgrounds and
Yang-Mills gauge theories in lower dimensions. These investigations,
triggered by Maldacena \cite{Maldacena}, include the conjecture that type
IIB string on an $AdS_{5}\times S^{5}$ background is dual to $N=4$, $d=4$
supersymmetric Yang-Mills theory. Since the isometry group of the manifold $%
AdS_{5}\times S^{5}$ is the superconformal group, the corresponding IIB
string theory could have a ghost-free, conformal supergravity theory as its
low energy limit. Therefore, it is interesting to revisit -- and extend --
the set of conformal supergravity theories. Here we find a alternative to
the classical conformal supergravity theories. The new theory has an action
linear in superconformal curvatures.

Intense activity in the 70s and early 80s provided what seemed to be a
complete picture of possible supergravity models. The demonstration in 1975
by Haag, \L opusa\'{n}ski and Sohnius \cite{Haag} of all possible
supersymmetries of the $S$ matrix showed clearly how supersymmetry could
overcome the limitations of the Coleman-Mandula theorem \cite{ColemanMandula}%
. It was only a short time before systematic classifications of the graded
Lie algebras emerged. The simple graded Lie algebras were classified by
Freund and Kaplansky \cite{FreundK}, and Kac \cite{Kac1}, \cite{Kac15} added
the exceptional algebras. Some work on classification was also provided in 
\cite{Sch1}, \cite{NahmSR}.

With these classifications available, Nahm \cite{Nahm} was able to identify
those graded algebras suitable for physical models in arbitrary spacetime
dimension by restricting to algebras with physical spin-statistics behavior,
compact internal symmetry, and an adjoint operation. Nahm went on to
determine the structure of all their flat space representations.

Simultaneously, other authors (\cite{Ferrara}-\cite{MacDowell}) explored
Poincar\'{e} and conformal supergravity theories based on the new symmetries
and developed the theory of supermanifolds. Of particular interest for our
purpose is the development of the group manifold (or group quotient) method
for constructing supermanifolds (see, eg., \cite{Neeman}-\cite{Brink} and in
particular, the review by Castellani, Fr\'{e} and van Nieuwenhuizen \cite
{CastFre}). The method described in \cite{CastFre} (who cite \cite{ReggeN};
see also \cite{Neeman}) is a modified version of techniques developed by
Cartan (for a complete treatment see Kobayashi and Nomizu \cite{Kobayashi}),
which has been generalized to supergroups. We now turn to a discussion of
conformal supergravity and examine the use of the group manifold method in
its construction.

The first comments on conformal supergravity by Freund (\cite{Freund1})
identify some of the properties of the superconformal gauge fields. The full
theory was then developed simultaneously and independently by Freund, Ferber
and Crispim-Romao in one series of articles (\cite{Freund1}, \cite{FF}, \cite
{CrispimRomao}, \cite{Ferber}) and by \thinspace Kaku, Townsend, van
Nieuwenhuizen and Ferrara (\cite{KTvN1}, \cite{FKTvN2}, \cite{vNTK3}, \cite
{KTVN4}, \cite{TvN2}) in another series of articles. One cannot doubt the
sense of excitement and urgency that accompanied these developments. Because
the review article, \cite{CastFre} uses similar methods to our own, we will
refer to the construction presented there.

The group manifold method provides a systematic way to implement a given
local symmetry as a gravity theory. Essentially, one begins with a Lie group
or graded Lie group, $G,$ containing the local symmetry (super)group of
interest, $H,$ as a sub-(super)group. Then the quotient $G/H$ is a manifold
with local $H$ symmetry. The cosets of $H$ in $G$ provide a projection from $%
G$ to $G/M,$ so the resulting structure is a principal fiber bundle with
fibers isomorphic to $H.$

In their implementation, Castellani, et al., independently choose the
dimension, $d$, of the final spacetime manifold. Their manifold may be any $%
d $-dimensional submanifold of the fiber bundle as long as $d\leq \dim
\left( G/H\right) $. They then introduce a connection one form, $h^{A},$ on
the bundle and write its curvature, $R^{A}$. The form of the curvature is
fully determined by the graded Lie algebra. Their construction is completed
by implementing two assumptions:

\begin{enumerate}
\item  The action is $H$-invariant integral of a $d$-form, 
\begin{equation}
S=\int \left( \Lambda +R^{A}v_{A}+R^{A}\wedge R^{B}v_{AB}+\cdots \right) 
\end{equation}

\item  The vacuum (which they define to be $R^{A}=0)$ is a solution of the
field equations.
\end{enumerate}

This last condition is necessary because of the arbitrary choice of the
spacetime dimension. Castellani, et al., must specify some condition of this
sort to fix the $H$-tensors $\Lambda ,v_{A},,v_{AB},\cdots .$ The so-called
``cohomology condition'' follows from the variation of the action when the
curvatures vanish, 
\begin{equation}
\frac{\delta }{\delta h^{A}}\Lambda +Dv_{A}=0
\end{equation}
This equation supplements the usual variational field equations. Solutions
to the combined cohomology and variational equations exist only for certain
subgroups $H$ and dimensions $d.$

Our construction begins with the same group quotient and fiber bundle
structure, but our subsequent assumptions differ in three ways. First, we do
not allow the choice of spacetime dimension in our construction. Instead, we
let the group and subgroup symmetries determine the dimension of the
physical spacetime by requiring $d=\dim (G/H).$ Second, we do not require
the physical space to be a submanifold of the group manifold. Rather, it may
be any manifold consistent with the local structure of the principal fiber
bundle $\left( G,G/H\right) $. Finally, we do not require vanishing
curvature to be a solution to the field equations. Indeed, we note that
vanishing curvature may be inconsistent with reduction to the $AdS$
background. Because of these last two differences, we are not required to
separately impose the cohomology condition.

As in \cite{CastFre}, the action may be any $H$-invariant $d$-form. However,
we note that since the action is built in an $H$-invariant way from the
curvatures, one cannot write an action before constructing the geometric
background.

In this way, the group structure is all that is required to construct a
general class of geometries with the desired symmetries, including dimension
of physical space, the expressions for curvatures of the connection, and the
relevant fields of the theory. From these curvatures it is straightforward
to write the most general linear action functional..

In summary, Castellani, Fr\'{e}, and van Nieuwenhuizen consider, in
principle, any subgroup $H$ of $G$ and any spacetime dimension consistent
with the cohomology and field equations, while we require manifold dimension 
$d=dim(G/H)$, drop the cohomology equations, and drop the constraint to zero
curvature solutions. Our method is likely more rigid and therefore more
predictive. In any case, demanding $d=dim(G/H)$ has interesting consequences
for the conformal and superconformal groups, and possibly for $M$ theory as
well. We now take a brief look at these theories, starting with the bosonic
case.

Gauging of the conformal group is implicit in any gauging of the
superconformal group, and thus the conformal supergravity theories
referenced above all contain conformal gaugings. A systematic presentation
of the possible gaugings that can be used to construct Poincar\'{e} and
conformal supergravity theories was provided by Ivanov and Niederle (\cite
{IvanovI},\cite{Ivanov}). Using group manifold methods, they reproduced the
gaugings present in the known conformal supergravity theories, and in \cite
{Ivanov} were the first to recognize an alternative gauging. The alternative
gauging sets the local symmety to the homothetic group, comprised of Lorentz
transformations and dilatations.

The use of the homothetic group as the local symmetry is not in itself a new
result. Indeed, the extension of the homothetic group by an internal
symmetry provides the residual bosonic symmetry of the conformal
supergravities considered in \cite{CastFre} and (\cite{Freund1}-\cite{TvN2}
). However, Ivanov and Niederle also set $d=\dim \left( G/H\right) =8,$
thereby introducing four new coordinates to the physical manifold. To deal
with the additional dimensions, they then resticted the $4$ new dimensions
to a submanifold generated by conformal transformations, thereby essentially
making these directions pure gauge degrees of freedom. The $4$ new
coordinates (or $n$ new coordinates for $n$-dim spacetimes) were freed from
this constraint in \cite{NCG}, using what is now called biconformal gauging
of the conformal group. The enlarged space still permits general relativity
on a Lorentzian submanifold. The extra dimensions participate as conjugate
(momentum-like) variables in a symplectic structure. Because the volume
element of the $8$- or $2n$-dim space is dimensionless, the biconformal
space allows actions linear in the curvature. Wehner and Wheeler \cite{WW}
showed that, with minimal or vanishing torsion, the most general linear
action is extremal only when there is a symplectic form and the Einstein
equation holds on a $4$- or $n$-dim submanifold. The remaining dimensions
may be identified (in all known classes of solutions) with coordinates on
the cotangent space.

The goal of the present work is to supersymmetrize $4$-dim gravity using the
alternative gauging. That is, we study $4$-dim biconformal supergravity.

\section{The Superconfornal graded Lie algebra}

The conformal group of a four dimensional spacetime is locally isomorphic to 
$O\left( 4,2\right) $. $Spin\left( 4,2\right) ,$ also locally isomorphic to $%
O\left( 4,2\right) ,$ gives a spinor representation for the conformal group.
Using the $4\times 4$ Dirac matrices, 
\begin{equation}
\left\{ \gamma ^{a},\gamma ^{b}\right\} =2\eta ^{ab}=2\ diag\left(
-1,1,1,1\right)  \label{Clifford}
\end{equation}
and defining 
\begin{eqnarray*}
\sigma ^{ab} &=&-\frac{1}{8}\left[ \gamma ^{a},\gamma ^{b}\right] \\
\gamma _{5} &=&i\gamma ^{0}\gamma ^{1}\gamma ^{2}\gamma ^{3}
\end{eqnarray*}
the full Clifford algebra has basis 
\[
\Gamma \in \left\{ 1,i1,\gamma ^{a},i\gamma ^{a},\sigma ^{ab},i\sigma
^{ab},\gamma _{5}\gamma ^{a},i\gamma _{5}\gamma ^{a},\gamma _{5},i\gamma
_{5}\right\} 
\]
where $a,b=0,1,2,3.$ We use this representation for the conformal sector of
the superconformal group.

The structure of the superconformal group, $SU(2,2|N)$ is well known (\cite
{Freund1}, \cite{FreundK}, \cite{Kac15}). To construct $SU(2,2|N)$ we demand
that the generators of the graded Lie algebra preserve a complex
super-metric, $\mathcal{H}$, diagonally composed of a \ $4$-$\dim $
Hermitian matix, $Q$ (Q is in fact a spinor metric-- see \cite{Crawford}),
and an $N$-$\dim $, anti-hermitian, (symmetric) matrix $P$, 
\begin{equation}
\mathcal{H}=\left( 
\begin{array}{cc}
Q &  \\ 
& P
\end{array}
\right) 
\end{equation}
The invariance condition is 
\begin{equation}
\mathcal{H}T+T^{\ddagger }\mathcal{H}=0
\end{equation}
where

\begin{equation}
T=\left( 
\begin{array}{cc}
A & B \\ 
C & D
\end{array}
\right) ,T^{\ddagger }=\left( 
\begin{array}{cc}
A^{\dagger } & -C^{\dagger } \\ 
B^{\dagger } & D^{\dagger }
\end{array}
\right)  \label{Transformation}
\end{equation}
where $T^{\ddagger }$ is the usual super-adjoint. \ We obtain, 
\begin{eqnarray}
QA+A^{\dagger }Q &=&0 \\
C &=&-P^{-1}B^{\dagger }Q \\
PD+D^{\dagger }P &=&0
\end{eqnarray}

We use a spinor representation for the conformal Lie algebra. Note that by
choosing, 
\begin{equation}
Q=-i\gamma ^{5}
\end{equation}
the invariance condition for the $A$-sector bosonic generators selects the
subset of Clifford generators 
\[
\Gamma _{0}\in \left\{ i1,\gamma ^{a},\sigma ^{ab},\gamma _{5}\gamma
^{a},\gamma _{5}\right\} 
\]
The last $15$ of these $16$ generators provide a manifest basis for the Lie
algebra $su\left( 2,2\right) .$

It is straightforward to find a representation of the Dirac matrices for
which $Q=diag\left( 1,1,-1,-1\right) ,$ and thus demonstrate that the 15
matrices listed above generate $SU\left( 2,2\right) .$ Thinking of $Q$ as a
Hermitian spinor metric, we are justified in calling the $Q$-invariant
subalgebra the isometry subalgebra of the Clifford algebra. Note that the
defining relationship of the Clifford algebra, eq.(\ref{Clifford}), is
invariant under $U\left( 4\right) $ transformations, and we may use this
freedom to select a real representation of the Dirac matrices. In a real
basis, $Q$ remains Hermitian but is necessarily antisymmetric, $Q=-Q^{t}.$
It follows that the generators of $SU\left( 2,2\right) $ are unitiarily
equivalent to a set preserving a symplectic form. 

\bigskip Choosing generators for the Lie algebra, we identify 
\begin{eqnarray}
M_{\quad b}^{a} &=&\eta _{bc}\sigma ^{ac}=-\frac{1}{8}\eta _{bc}\left[
\gamma ^{a},\gamma ^{c}\right]  \label{M} \\
P_{a} &=&\frac{1}{2}\eta _{ab}\left( 1+\gamma _{5}\right) \gamma ^{b}=\frac{1%
}{2}\eta _{ab}\gamma ^{b}\left( 1-\gamma _{5}\right)  \label{P} \\
K^{a} &=&\frac{1}{2}\left( 1-\gamma _{5}\right) \gamma ^{a}=\frac{1}{2}
\gamma ^{a}\left( 1+\gamma _{5}\right)  \label{K} \\
D &=&-\frac{1}{2}\gamma _{5}  \label{D}
\end{eqnarray}
and compute the Lie algebra, which is listed in Appendix 1. Here, $%
M_{ab}=-M_{ba}=\eta _{ac}M_{b}^{c}$ are the Lorentz rotation generators, $%
P_{a}$ the translations, $K^{a}$ the special conformal transformations, and
D the dilatation. With modified sign conventions, these agree with the work
of \cite{KTvN1} and \cite{FKTvN2}.

In addition to giving a symplectic representation, a real representation for
the Dirac matrices is a convenience in writing real-valued action
functionals. With real Dirac matrices, $M^{a}{}_{b}$ is real, $D$ is pure
imaginary, and $P_{a}$ , $K^{a}$ are complex conjugates of one another.

Returning to the invariance conditions, we note that the $D$-sector
generators preserve the anti-Hermitian form $P,$ which in the real
representation is symmetric, $P^{t}=P.$ Therefore, the internal symmetry is $%
U\left( N\right) .$ The set of generators includes $N(N-1)/2$ real,
antisymmetric generators and $N(N+1)/2$ imaginary, symmetric generators to
constitute the required $N^{2}$ generators.

Among the bosonic symmetries are two commuting generators: $i1_{4}$ in the $A
$-sector and $i1_{N}$ in the $D$-sector, where $1_{4}$ and $1_{N}$ denote
the $4$- and $N$-dim identies, respectively. By demanding vanishing
superdeterminant for the superconformal group, we eliminate these in favor
of the supertraceless combination 
\[
E=-\frac{i}{4}\left( 
\begin{array}{cc}
1_{4} &  \\ 
& \frac{4}{N}1_{N}
\end{array}
\right) 
\]
The generator $E$ functions as a central charge in the conformal subalgebra
and appears in one fermionic anticommutator.

There are therefore a total of $N^{2}+16$ bosonic generators and $8N$
fermionic generators. Explicit forms for all of the generators (and the
complete $su(2,2|N)$ algebra) are presented in Appendix 1.

\section{Maurer Cartan Structure Equations:}

We now define the set of super differential forms dual to the generators of
the super Lie algebra. In general, the dual differential forms are defined
as follows. 
\[
\left\langle G_{\Sigma },\mathbf{\omega }^{\Pi }\right\rangle \equiv \delta
_{\Sigma }^{\Pi } 
\]
Here the indices $\Pi ,\Sigma $ run over the types of indices present in the
Lie algebra and differential forms are bold. $G_{\Sigma }$ represents an
arbitrary generator of the super Lie algebra and\textbf{\ }$\mathbf{\omega }%
^{\Pi }$ is the corresponding Lie algebra valued one form. We utilize
differential forms in order to make the expressions for the action more
manageable (\cite{CastFre}). Note that by definition, we are assigning the
differential forms,\textbf{\ }$\mathbf{\omega }_{\Lambda },$ to have the
opposite conformal weight of their corresponding generators, $G^{\Sigma }.$
Explicitly we define the form $\mathbf{\omega }^{\Pi }\in \left\{ \mathbf{%
\omega }_{b}^{a},\mathbf{\omega }_{a},\mathbf{\omega }^{a},\mathbf{\omega },%
\mathbf{\psi }_{\beta }^{B},\mathbf{\chi }_{\beta }^{B},\mathbf{\alpha },%
\mathbf{\pi }_{R}^{\rho \sigma },\mathbf{\pi }_{I}^{\rho \sigma }\right\} $
by: 
\begin{equation}
\begin{array}{cc}
\left\langle M_{\quad b}^{a},\mathbf{\omega }_{d}^{c}\right\rangle =\delta
_{d}^{a}\delta _{b}^{c}-\eta ^{ac}\eta _{bd} & \left\langle G_{A}^{\alpha +},%
\mathbf{\chi }_{\beta }^{B}\right\rangle =\delta _{\beta }^{\alpha }\delta
_{A}^{B} \\ 
\left\langle P_{a},\mathbf{\omega }^{b}\right\rangle =\delta _{a}^{b} & 
\left\langle G_{A}^{\alpha -},\mathbf{\psi }_{\beta }^{B}\right\rangle
=\delta _{\beta }^{\alpha }\delta _{A}^{B} \\ 
\left\langle K^{a},\mathbf{\omega }_{b}\right\rangle =\delta _{b}^{a} & 
\left\langle D_{R}^{\mu \nu },\mathbf{\pi }_{\alpha \beta }^{R}\right\rangle
=\delta _{\alpha }^{\mu }\delta _{\beta }^{\nu } \\ 
\left\langle D,\mathbf{\omega }\right\rangle =1 & \left\langle D_{I}^{\mu
\nu },\mathbf{\pi }_{\alpha \beta }^{I}\right\rangle =\delta _{\alpha }^{\mu
}\delta _{\beta }^{\nu } \\ 
\left\langle E,\mathbf{\alpha }\right\rangle =1 & 
\end{array}
\end{equation}
The first four forms listed on the left are associated with the Lorentz,
translation, co-translation, and dilatation generators, respectively. We
refer to\textbf{\ } $\mathbf{\omega }_{b}^{a}$ as the spin-connection, $%
\mathbf{\omega }^{a}$ as the solder form, $\mathbf{\omega }_{a}$ as the
co-solder form, and $\mathbf{\omega }$ as the Weyl vector. $\mathbf{\psi }%
_{\beta }^{B} $ and $\mathbf{\chi }_{\beta }^{B}$ correspond to the
fermionic generators while the remaining three one-forms, $\mathbf{\alpha
,\pi }_{R}^{\rho \sigma },\mathbf{\pi }_{I}^{\rho \sigma }$ are associated
with the internal symmetry.

The Maurer-Cartan structure equations are defined in general by 
\begin{equation}
0=\mathbf{d\omega }^{\Sigma }+\frac{1}{2}c_{\Gamma \Delta }{}^{\Sigma }%
\mathbf{\omega }^{\Gamma }\mathbf{\omega }^{\Delta }
\end{equation}
and are fully equivalent to the Lie algebra relations, with $\mathbf{d}^{2}=0
$ providing the Jacobi identies. Here $c_{\Gamma \Delta }{}^{\Sigma }$ are
the structure constants of the graded Lie algebra and the standard wedge
product is assumed between all differential forms. Using the group quotient
method (\cite{Kobayashi}, \cite{Neeman}, \cite{ReggeN}, \cite{CastFre}), the
connection is generalized, giving the curvature 2-forms $\mathbf{\Omega }^{%
\Sigma },$ 
\begin{equation}
\mathbf{\Omega }^{\Sigma }=\mathbf{d\omega }^{\Sigma }+\frac{1}{2}c_{\Gamma 
\Delta }{}^{\Sigma }\mathbf{\omega }^{\Gamma }\mathbf{\omega }^{\Delta }
\label{Cartan}
\end{equation}
according to the Cartan structure equations. When we form the group quotient
between $SU(2,2|N)$ and our chosen isotropy subgroup, the curvature 2-forms
are required to be \textit{horizontal}, that is, expandable only in the
basis forms spanning the co-tangent space to the quotient manifold. For
example, the usual Riemannian structure of general relativity arises from
the quotient of the Poincar\'{e} group by its Lorentz subgroup. The dual
forms for the Poincar\'{e} group are of two types: the spin connection $%
\mathbf{\omega }_{b}^{a}$ and the solder form $\mathbf{e}^{a}.$ The
horizontal curvatures may be expanded bilinearly in the solder forms only, $%
\mathbf{R}_{b}^{a}=\frac{1}{2}\mathbf{R}_{bcd}^{a}\mathbf{e}^{c}\mathbf{e}%
^{d}.$

Regardless of the group quotient we choose, the general form of the
curvatures is the same until they are expanded in basis forms. Thus, we may
immediately write the following expressions for the curvature 2-forms using
the structure constants from Appendix 1. We have 
\begin{eqnarray}
\mathbf{\Omega }_{b}^{a} &=&\mathbf{d\omega }_{b}^{a}-\mathbf{\omega }%
_{b}^{c}\mathbf{\omega }_{c}^{a}-2\mathbf{\omega }_{b}\mathbf{\omega }%
^{a}+2\eta ^{ac}\eta _{bd}\mathbf{\omega }_{c}\mathbf{\omega }^{d}-P^{\alpha
\beta }\left[ \sigma _{\quad b}^{a}\right] _{AB}\mathbf{\chi }_{\alpha }^{A}%
\mathbf{\psi }_{\beta }^{B}  \label{Curvature} \\
\mathbf{\Omega }^{a} &=&\mathbf{d\omega }^{a}-\mathbf{\omega }^{c}\mathbf{%
\omega }_{c}^{a}-\mathbf{\omega \omega }^{a}+\frac{1}{2}P^{\alpha \beta
}\left[ \gamma ^{a}\right] _{\left( AB\right) }\mathbf{\psi }_{\alpha }^{A}%
\mathbf{\psi }_{\beta }^{B}  \label{Torsion} \\
\mathbf{\Omega }_{a} &=&\mathbf{d\omega }_{a}-\mathbf{\omega }_{a}^{c}%
\mathbf{\omega }_{c}-\mathbf{\omega }_{a}\mathbf{\omega }+\frac{1}{2}%
P^{\alpha \beta }\left[ \gamma _{a}\right] _{\left( AB\right) }\mathbf{\chi }%
_{\alpha }^{A}\mathbf{\chi }_{\beta }^{B}  \label{Co-torsion} \\
\mathbf{\Omega } &=&\mathbf{d\omega -}2\mathbf{\omega }^{a}\mathbf{\omega }%
_{a}-\frac{1}{2}P^{\alpha \beta }Q_{AB}\mathbf{\chi }_{\alpha }^{A}\mathbf{\
\psi }_{\beta }^{B}  \label{Dilatation}
\end{eqnarray}
for the usual supersymmetric generalization of the bosonic curvatures, and 
\begin{eqnarray}
\mathbf{\Theta }_{\beta }^{B} &=&\mathbf{d\psi }_{\beta }^{B}-\left[ \frac{1%
}{2}\mathbf{\omega }_{a}^{b}\sigma _{\quad b}^{a}\right] _{\quad A}^{B}%
\mathbf{\psi }_{\beta }^{A}+\left[ \mathbf{\omega }^{a}\gamma _{a}\right]
_{\quad A}^{B}\mathbf{\chi }_{\beta }^{A}-\frac{1}{2}\mathbf{\omega \psi }%
_{\beta }^{B}  \nonumber \\
&&-2iP^{\alpha \mu }\mathbf{\pi }_{\mu \beta }^{R}\mathbf{\psi }_{\alpha
}^{B}-2P^{\alpha \mu }\mathbf{\pi }_{\mu \beta }^{I}\left[ \gamma
_{5}\right] _{\quad A}^{B}\mathbf{\psi }_{\alpha }^{A}  \label{PsiCurvature}
\\
\overline{\mathbf{\Theta }}_{\beta }^{B} &=&\mathbf{d\chi }_{\beta
}^{B}-\left[ \frac{1}{2}\mathbf{\omega }_{a}^{b}\sigma _{\quad b}^{a}\right]
_{\quad A}^{B}\mathbf{\chi }_{\beta }^{A}+\left[ \mathbf{\omega }_{a}\gamma
^{a}\right] _{\quad A}^{B}\mathbf{\psi }_{\beta }^{A}+\frac{1}{2}\mathbf{%
\omega \chi }_{\beta }^{B}  \nonumber \\
&&-2iP^{\alpha \mu }\mathbf{\pi }_{\mu \beta }^{R}\mathbf{\chi }_{\alpha
}^{B}+2P^{\alpha \mu }\mathbf{\pi }_{\mu \beta }^{I}\left[ \gamma
_{5}\right] _{\quad A}^{B}\mathbf{\ \chi }_{\alpha }^{A}
\label{ChiCurvature} \\
\mathbf{\Pi }_{\rho \sigma }^{R} &=&\mathbf{d\pi }_{\rho \sigma
}^{R}+2iP^{\beta \mu }\left( \mathbf{\pi }_{\rho \beta }^{R}\mathbf{\pi }%
_{\mu \sigma }^{R}-\mathbf{\pi }_{\rho \beta }^{I}\mathbf{\pi }_{\mu \sigma
}^{I}\right) -\frac{i}{4}Q_{AB}\left( \mathbf{\chi }_{\rho }^{A}\mathbf{\psi 
}_{\sigma }^{B}-\mathbf{\chi }_{\sigma }^{A}\mathbf{\psi }_{\rho }^{B}\right)
\label{InternalReal} \\
\mathbf{\Pi }_{\rho \sigma }^{I} &=&\mathbf{d\pi }_{\rho \sigma
}^{I}+2iP^{\beta \mu }\left( \mathbf{\pi }_{\rho \beta }^{R}\mathbf{\pi }%
_{\mu \sigma }^{I}+\mathbf{\pi }_{\sigma \beta }^{R}\mathbf{\pi }_{\mu \rho
}^{I}\right) \\
&&+\frac{1}{4}\left[ \gamma _{5}\right] _{AB}\left( \mathbf{\chi }_{\rho
}^{A}\mathbf{\psi }_{\sigma }^{B}+\mathbf{\chi }_{\sigma }^{A}\mathbf{\ \psi 
}_{\rho }^{B}\right)  \label{InternalImag} \\
\mathbf{A} &=&\mathbf{d\alpha }+\frac{1}{2}\left( \left( \gamma _{5}\right)
_{B}^{C}Q_{CA}\mathbf{\chi }_{\alpha }^{A}\mathbf{\psi }_{\beta
}^{B}P^{\alpha \beta }+\left( \gamma _{5}\right) _{A}^{C}Q_{CB}\mathbf{\psi }%
_{\alpha }^{A}\mathbf{\chi }_{\beta }^{B}P^{\alpha \beta }\right)
\label{Central}
\end{eqnarray}
for the fermionic and internal curvatures.

We define $\mathbf{\Theta }_{\beta }^{B}$ to be the fermionic curvature and $%
\mathbf{\Pi }_{\rho \sigma }^{R},\mathbf{\Pi }_{\rho \sigma }^{I},\mathbf{A}$
to be the internal symmetry curvatures ($\mathbf{\Pi }_{\rho \sigma }^{R}, 
\mathbf{\Pi }_{\rho \sigma }^{I}$ refer to the real and imaginary
components, respectively, of the internal symmetry curvatures). Note that in
the equations listed above $\mathbf{\omega }^{a},\mathbf{\omega }_{a}$ refer
to two independent one forms associated with distinct generators. The
position of the lower case Latin indices is used to designate the generator,
it does \textit{not} refer to any use of the metric.

It is important to recognize that eqs.(\ref{Curvature}-\ref{Central}) do not
yet fully define the curvatures because we have not yet specified the
subgroup which determines horizontality. Knowing this group determines not
only the cotangent basis forms in which they are to be expanded, but also
how these component curvatures mix under the residual fiber (gauge)
symmetry. We now turn to these questions.

\section{Gauging the Supergroup}

In general, to gauge the group, an isotropy subgroup (any subgroup
containing no subgroup normal in the full group) is chosen. In keeping with
our comments in the introduction, this choice determines the dimension and
local character of the physical superspace. The quotient of the full group
by the isotropy subgroup is a manifold whose dimension is given by the
difference in the dimensions between the full group and the isotropy
subgroup. The curvatures of the manifold follow by generalizing the
connection $\mathbf{\omega }^{\Sigma }$ and demanding horizontality. We
allow any global structure consistent with this local structure.

Much is known about the bosonic case of conformal gauging. As described in 
\cite{Ivanov}, \cite{IvanovI}, and \cite{NCG}, the demand that the local
symmetry contain both Lorentz transformations and dilatations leaves only
two choices for the quotient. The first of these is to take the quotient of
the conformal group by the subgroup built from Lorentz transformations,
dilatations and special conformal transformations (i.e., the inhomogeneous
homothetic group). The quotient manifold is then $4$-dimensional, and is
immediately identified with spacetime. The alternative \textit{biconformal}
gauging takes the quotient of the conformal group by the homogeneous
homothetic group, consisting of Lorentz transformations and dilatations only.

The first case has been treated abundantly in the literature, as discussed
in the introduction. While the homothetic group is used for the isotropy
subgroup in \cite{CastFre}, the manifold is still taken to be $4$%
-dimensional and the cohomology equations are imposed by hand.

By contrast, the biconformal gauging of the conformal group of a
compactified, $n$-dim spacetime produces a $2n$-dim space. This space is
spanned by $n$ coordinates with units of length and another $n$ coordinates
with units of inverse length. As a result, the volume form is dimensionless
and it is possible to write a gravity action which is linear in the
curvature. Assuming minimal \cite{WW} or vanishing \cite{NCG} torsion, the
resulting field equations reduce in a particular subset of conformal gauges
to the Einstein equation of general relativity on an $n$-dimensional
submanifold. In the final symmetry of the space, the translational and
special conformal symmetries become general coordinate symmetry on the base
manifold, leaving the Lorentz and dilatational curvatures as local
symmetries. The extra $n$ dimensions participate in a symplectic structure
that has been shown to be consistent with the Hamiltonian dynamics \cite
{GaugeNewton} of an $n$-dim configuration space.

Our central aim is now to reproduce this result for a supersymmetric
extension of the conformal group and write linear actions over the resulting
space.

In order to define a fiber bundle over a supermanifold we must generalize
the isotropy subgroup, the homogeneous homothetic group, to a sub-supergroup 
$H$ of the entire supergroup. We demand two properties of $H$:

\begin{enumerate}
\item  The $A$-sector of the bosonic part of $H$ must consist of Lorentz
transformations and dilatations.

\item  The $D$-sector of the bosonic part of $H$ must be $U\left( N\right) ,$
thereby retaining the entire $D$-sector as a local internal symmetry.
\end{enumerate}

We now show that there are three possible choices for $H$ satisfying these
conditions. Two of these are mathematically equivalent, so there are two
distinct group quotients that give rise to the desired homothetic bosonic
fiber symmetry. The proof is as follows.

First consider condition $1.$ The homothetic algebra may be characterized as
the dilatationally invariant subalgebra of the conformal algebra. That is,
the homothetic generators, $W$, are exactly those that satisfy, 
\begin{eqnarray}
\left[ \gamma _{5},W\right] &=&0  \label{Homothetic} \\
QW+W^{\dagger }Q &=&0  \label{Conformal}
\end{eqnarray}
Alternatively, since all elements of the graded Lie algebra already satisfy
eq.(\ref{Conformal}), we may say that the subalgebra that preserves 
\[
\alpha Q+\beta Q\gamma _{5} 
\]
for any fixed $\alpha ,\beta \neq 0,$ is homothetic.

Next, we generalize this condition to an arbitrary element of $su(2,2|N)$
and impose both conditions 1 and 2. Thus, the subset of $su(2,2|N)$
generators leaving any matrix, $\mathcal{M}$, of the form 
\[
\mathcal{M}=\left( 
\begin{array}{cc}
\alpha Q+\beta Q\gamma _{5} & R \\ 
S & J
\end{array}
\right) 
\]
invariant generates a subgroup of the superconformal group$.$ We demand this
subgroup to be our isotropy, $H.$ Letting $T$ be as in eq.(\ref
{Transformation}), invariance of $\mathcal{M}$, namely, $\mathcal{M}%
T+T^{\ddagger }\mathcal{M}=0$ for all $T$ leads directly to the conditions $
R=S=0$ and $J=\lambda P.$ The three possible solutions depend on whether 
\[
\det \left( \left( \alpha -\lambda \right) Q+\beta Q\gamma _{5}\right) 
\]
vanishes or not (see Appendix II). The three possibilities are:

\begin{enumerate}
\item  $\left( \alpha -\lambda \right) \neq \beta .$ No nonzero spinor $B$
survives in the local symmetry. All fermionic degrees of freedom are then
coordinate degrees of freedom, so we have a 16-dim superspace spanned by $%
\mathbf{\omega }^{a},$ $\mathbf{\omega }_{a},\mathbf{\chi }$ and $\mathbf{%
\psi }$ with local homothetic and $U(N)$ symmetry. The supervolume element
is dimensionless.

\item  $\left( \alpha -\lambda \right) =\beta \neq 0.$ Only left handed
spinors $B$ provide local symmetries. In this case, each local fermionic
symmetry $B$ must satisfy: 
\[
\left( 1+\gamma _{5}\right) B=0
\]
Therefore, half of the fermionic degrees of freedom (those generated by $%
G_{A}^{+})$ lie on the fiber and half become coordinate degrees of freedom.
The volume element, $\sim \mathbf{\chi }^{1}\mathbf{\chi }^{2}\mathbf{\chi }
^{3}\mathbf{\chi }^{4},$ has scaling dimension $(length)^{2N}$ and, since $%
\mathbf{\psi }$ and $\mathbf{\chi }$ are complex conjugates, there is no
evident real-valued action linear in the curvatures.

\item  $\left( \alpha -\lambda \right) =-\beta \neq 0.$ Only right handed
spinors $B$ provide local symmetry The local fermionic symmetries satisfy 
\[
\left( 1-\gamma _{5}\right) B=0
\]
so they are those generated by $G_{A}^{-}.$ Again, the fermionic degrees of
freedom are split between coordinate and fiber, giving a volume element $%
\sim $ $\mathbf{\psi }^{1}\mathbf{\psi }^{2}\mathbf{\psi }^{3}\mathbf{\psi }%
^{4},$ of scaling dimension $(length)^{-2N}.$ There is no real-valued action
linear in the curvatures.
\end{enumerate}

\smallskip

The two chiral solutions may ultimately prove to be of interest, since they
display heteroticity, but they will not occupy us further here. The first
case, in which all fermionic degrees of freedom are realized as superspace
coordinates, has the dimensionless volume form characteristic of biconformal
gauging. We now examine this case in detail.

First, to accomplish the group quotient, we note that both $Q$ and $Q\gamma
_{5}$ are independent invariants. Then the isotropy subgroup is generated by
those transformations leaving both the super-metric, $\mathcal{H},$ and 
\begin{equation}
\mathcal{M}=\left( 
\begin{array}{cc}
Q\gamma _{5} & 0 \\ 
0 & 0
\end{array}
\right) 
\end{equation}
invariant. This amounts to all Lorentz transformations, dilatations, the
central charge $E,$ and $U(N)$ transformations. These span a $8+N^{2}$
dimensional submanifold of $SU(2,2|N),$ while the quotient $G/H$ is $16$%
-dimensional.

Next, we implement the requirement for horizontal curvatures. Each curvature
is now defined to be bilinear in the set of basis forms 
\[
\mathbf{\omega }^{\Pi }\in \left\{ \mathbf{\omega }_{a},\mathbf{\omega }^{a},%
\mathbf{\psi }_{\beta }^{B},\mathbf{\chi }_{\beta }^{B},\right\} 
\]
Thus, for each curvature, 
\[
\mathbf{\Omega }^{\Sigma }=\frac{1}{2}\Omega ^{\Sigma }{}_{\Pi \Lambda }%
\mathbf{\omega }^{\Pi }\mathbf{\omega }^{\Lambda } 
\]
Expanding explicitly, we adopt the following notational conventions for each
curvature: 
\begin{eqnarray}
\mathbf{\Omega }^{\Sigma } &=&\frac{1}{2}R_{\quad ab}^{\Sigma }\mathbf{\
\omega }^{a}\mathbf{\omega }^{b}+R_{\quad b}^{\Sigma a}\mathbf{\omega }_{a}%
\mathbf{\omega }^{b}+\frac{1}{2}R^{\Sigma ab}\mathbf{\omega }_{a}\mathbf{%
\omega }_{b}  \nonumber \\
&&+R_{\quad a\widetilde{B}}^{\Sigma }\mathbf{\omega }^{a}\mathbf{\psi }^{%
\widetilde{B}}+\bar{R}_{\quad aA}^{\Sigma }\mathbf{\omega }^{a}\mathbf{\chi }%
^{A}  \nonumber \\
&&+R_{\quad \tilde{A}}^{\Sigma a}\mathbf{\omega }_{a}\mathbf{\psi }^{%
\widetilde{A}}+R_{\quad A}^{\Sigma a}\mathbf{\omega }_{a}\mathbf{\chi }^{A} 
\nonumber \\
&&+\frac{1}{2}R_{\quad \widetilde{A}\widetilde{B}}^{\Sigma }\mathbf{\psi }^{%
\widetilde{A}}\mathbf{\psi }^{\widetilde{B}}+R_{\quad A\widetilde{B}%
}^{\Sigma }\mathbf{\chi }^{A}\mathbf{\psi }^{\widetilde{B}}+\frac{1}{2}\bar{R%
}_{\quad AB}^{\Sigma }\mathbf{\chi }^{A}\mathbf{\chi }^{B}
\end{eqnarray}
where indices with tildes are contracted with $\mathbf{\psi }^{\widetilde{A}%
} $ as opposed to $\mathbf{\chi }^{A}.$

Finally, each connection form and curvature is now regarded as an $H$-tensor
rather than a superconformal tensor. This means that eqs.(\ref{Curvature}-%
\ref{Central}) now represent nine independent $H$-tensors instead of a
single $su(2,2|N)$ tensor. Indeed, even the ten components $R_{\quad ab}^{%
\Sigma },\ldots ,\bar{R}_{\quad AB}^{\Sigma }$ of each have no mixing under
homothetic transformations, and therefore provide a total of $90$
independent tensor fields.

The basis forms $\mathbf{\omega }_{a},\mathbf{\omega }^{b},\mathbf{\chi }%
^{A},\mathbf{\psi }^{B}$ all transform tensorially, under the homothetic
gauge group, as do the curvatures, $\mathbf{\Omega }_{a}.\mathbf{\Omega }%
^{a},\mathbf{\Theta }^{A},\overline{\mathbf{\Theta }}^{A}$ and $\mathbf{%
\Omega }_{b}^{a}.$ The remaining curvatures, $\mathbf{\Omega }$ and $\mathbf{%
A}$, are invariant under the gauge transformations. Thus, the supersymmetry
of the space is entirely in the coordinate transformations of the superspace
formulation. General coordinate transformations on the base manifold can mix
the components of the fermionic and bosonic basis forms.

These gauge transformations differ markedly from those of (\cite{Freund1}-%
\cite{TvN2}), since these theories retain local Lorentz, dilatational and
special conformal symmetry, together with the fermionic transformations. In 
\cite{CastFre}, the bosonic local symmetry is reduced to Lorentz and
dilatational, plus supersymmetries. In the present formalism, the full local
symmetry is simply Lorentz and dilatational, while the fermionic
transformations become coordinate transformations of superspace. Thus, our
construction results in far more invariant quantities, and makes it possible
to easily write purely geometric invariant actions.

Below, we describe actions for the biconformal gauging of $SU(2,2|N).$ We
present the most general, gauge invariant, linear actions of both even and
odd parity. The even parity action gives rise to general relativity in the
bosonic sector of the superspace, and we show that the Rarita-Schwinger
equation is contained in the fermionic sector.

\section{Biconformal Actions}

Before addressing possible actions, we must consider the volume element for
super-biconformal space. Since the bosonic portion of the base manifold is
spanned by the solder and co-solder forms we will first define, 
\[
\mathbf{\phi }_{bosonic}=\mathbf{\phi }_{b}=\varepsilon _{acde}\varepsilon
^{bfgh}\mathbf{\omega }_{bfgh}\mathbf{\omega }^{acde} 
\]
where 
\[
\mathbf{\omega }_{bfgh}=\mathbf{\omega }_{b}\mathbf{\omega }_{f}\mathbf{%
\omega }_{g}\mathbf{\omega }_{h} 
\]
and $\varepsilon _{acde}$ is the four dimensional Levi-Civita tensor. Again,
the mixed index position indicates the scaling weight of the indices and not
any use of the metric. A full derivation and justification of this volume
element is given in \cite{WW}\textbf{. \ }It is important to note that the
real-valued $8$-form $\mathbf{\phi }_{b}$ is both dilatationally and Lorentz
invariant. The Levi-Civita tensor is normalized such that traces are given
by 
\[
\varepsilon _{a_{1}...a_{p}c_{p+1...}c_{n}}\varepsilon
^{b_{1}...b_{p}c_{p+1}...c_{n}}=p!\left( n-p\right) !\delta
_{a_{1}...a_{p}}^{b_{1}...b_{p}} 
\]
where $\delta $ represents the following totally antisymmetric tensor, 
\[
\delta _{a_{1}...a_{p}}^{b_{1}...b_{p}}\equiv \delta
_{a_{1}...a_{p}}^{\left[ b_{1}...b_{p}\right] } 
\]

The fermionic portion of the base manifold is spanned by the spinor-valued
one forms, $\mathbf{\psi }$ and $\mathbf{\chi .}$ The volume form is
therefore proportional to 
\[
\mathbf{\psi }^{1}\mathbf{\psi }^{2}\mathbf{\psi }^{3}\mathbf{\psi }^{4}%
\mathbf{\chi }^{1}\mathbf{\chi }^{2}\mathbf{\chi }^{3}\mathbf{\chi }^{4} 
\]
However, it is desirable to write a manifestly tensorial expression. This
makes subsequent calculations simpler, but there is a difficulty in
constructing one. Unlike bosonic differential forms, fermionic forms commute
so the wedge product does not automatically eliminate quadratic terms such
as $\mathbf{\psi }^{1}\wedge \mathbf{\psi }^{1}.$ However, this difficulty
is readily overcome by noting the following ideas.

Differential $p$-forms may be defined as maps from $p$-dimensional volumes
into the reals. Thus, for example, the $1$-form $\mathbf{f}=f(x)\mathbf{d}x$
maps $\mathbf{f:}C\rightarrow R$ according to $r=\int f(x)\mathbf{d}x.$ This
is a useful point of view for fermionic forms. By the rules of Berezin
integration (\cite{Berezin}, \cite{Rogers2}, \cite{Rabin2}), integrals over
a pair of identical fermionic forms vanish, regardless of the integrand.
Specifically, although $\mathbf{d}\theta \mathbf{d}\theta $ is not
manifestly zero by symmetry, it vanishes on every complete superspace
integral. Ignoring the obvoius ambiguities, consider what map $\mathbf{d}
\theta \mathbf{d}\theta $ must be: 
\begin{eqnarray*}
\int \int f\left( \theta \right) \mathbf{d}\theta \mathbf{d}\theta &=&\int
\int \left( a+b\theta \right) \mathbf{d}\theta \mathbf{d}\theta \\
&=&\int \left( \int \left( a+b\theta \right) \mathbf{d}\theta \right) 
\mathbf{d}\theta \\
&=&\int b\mathbf{d}\theta \\
&=&0
\end{eqnarray*}
Alternatively, we note that in order for two $\mathbf{d}\theta $ integrals
to fail to vanish, we would require two factors of $\theta $ in the
integrand. This also vanishes. Therefore, we are justified in defining
quadratic terms such as $\mathbf{\psi }^{1}\wedge \mathbf{\psi }^{1}$ to be
equivalent to the zero map, and therefore zero.

For two fermionic degrees of freedom, $\theta _{1}$ and $\theta _{2},$ we
may therefore write 
\begin{eqnarray*}
\frac{1}{2}\left( \mathbf{d}\theta _{1}+\mathbf{d}\theta _{2}\right) ^{2} &=&%
\frac{1}{2}\mathbf{d}\theta _{1}\mathbf{d}\theta _{1}+\mathbf{d}\theta _{1}%
\mathbf{d}\theta _{2}+\frac{1}{2}\mathbf{d}\theta _{2}\mathbf{d}\theta _{2}
\\
&\cong &\mathbf{d}\theta _{1}\mathbf{d}\theta _{2}
\end{eqnarray*}
where we use the equivalence of quadratics to zero in the last step. With
this convention in mind, any purely eighth order polynomial in $\mathbf{\psi 
}$ and $\mathbf{\chi }$ is proportional to the $N=1$ volume element. A
simple covariant expression for the volume form is therefore 
\[
\mathbf{\phi }_{f}=\frac{1}{4!}\left( Q_{AB}\mathbf{\chi }^{A}\mathbf{\psi }%
^{B}\right) ^{4} 
\]
It is straightforward to check that this reduces to $\mathbf{\psi }^{1}%
\mathbf{\psi }^{2}\mathbf{\psi }^{3}\mathbf{\psi }^{4}\mathbf{\chi }^{1}%
\mathbf{\chi }^{2}\mathbf{\chi }^{3}\mathbf{\chi }^{4}.$ The generalization
to arbitrary $N$ is immediate: 
\[
\mathbf{\phi }_{f}=\left( Q_{AB}P^{\alpha \beta }\mathbf{\chi }_{\alpha }^{A}%
\mathbf{\ \psi }_{\beta }^{B}\right) ^{4N} 
\]
Finally, the full volume form over the superspace is 
\[
\mathbf{\Phi }=\mathbf{\phi }_{b}\mathbf{\phi }_{f} 
\]

To construct an action linear in the eight curvatures, 
\[
\left\{ \mathbf{\Omega }_{b}^{a},\mathbf{\Omega }^{a},\mathbf{\Omega }_{a},%
\mathbf{\Omega },\mathbf{A},\mathbf{\Pi }_{\alpha \beta }^{R},\mathbf{\Pi }%
_{\alpha \beta }^{I},\mathbf{\Theta }_{\beta }^{B},\overline{\Theta }_{\beta
}^{B}\right\} 
\]
we first note the additional available tensors fields. These include the
Dirac matrices, 
\[
\left\{ \gamma ^{a},\gamma _{5},\sigma _{b}^{a},\gamma _{5}\gamma
^{a}\right\} 
\]
together with the set consisting of the Minkowski metric, $\eta _{ab},$ the
spinor metric, $Q_{AB,}$ the $U(N)$ metric, $P^{\alpha \beta },$ and the
Levi-Civita tensor, $\varepsilon _{abcd}$. We define the Dirac matrices to
be of zero conformal weight and covariantly constant, $D_{\Sigma }\Gamma
_{0}=0.$

We next use the following fact to construct the action. Let $\Phi _{\Sigma }$
be a general tensor-valued $8\left( N+1\right) -2$ form with index $\Sigma $
of arbitrary type and let $\mathbf{\Omega }^{\Sigma }$ be any curvature
2-form. Then their product, 
\[
\mathbf{\Omega }^{\Sigma }\Phi _{\Sigma } 
\]
must be proportional to a complete volume form, 
\[
\mathbf{\Omega }^{\Sigma }\Phi _{\Sigma }=\Omega ^{\Sigma \Lambda }{}_{\Pi
\Delta }S_{\Sigma \Lambda }{}^{\Pi \Delta }\Phi 
\]
where $S_{\Sigma \Lambda }{}^{\Pi \Delta }$ is built from the available
tensor fields characterized above.

The most general, even parity, homothetic gauge-invariant action linear in
the curvatures for the case of a $U(N)$ internal symmetry is 
\begin{eqnarray}
S &=&\int \left\{ \left( \alpha _{1}\mathbf{\Omega }_{b}^{a}\left[ \sigma
_{a}^{b}\right] _{AB}+\left( \alpha _{2}\mathbf{\Omega }+\alpha _{3}\mathbf{A%
}\right) Q_{AB}\right) \mathbf{\chi }_{\alpha }^{A}\mathbf{\psi }_{\beta
}^{B}P^{\alpha \beta }\left( \mathbf{\chi \psi }\right) ^{2}\mathbf{\phi }%
_{b}\right.  \nonumber \\
&&+\left( \alpha _{4}\mathbf{\Omega }_{b}^{a}+\left( \alpha _{5}\mathbf{\
\Omega }+\alpha _{6}\mathbf{A}\right) \delta _{b}^{a}\right) \varepsilon
_{acde}\varepsilon ^{bfgh}\mathbf{\omega }_{fgh}\mathbf{\omega }^{cde}%
\mathbf{\phi }_{f}  \nonumber \\
&&+\alpha _{7}(\mathbf{\Omega }^{m}\left[ \gamma _{m}\right] _{BD}\mathbf{\
\chi }_{\alpha }^{B}\mathbf{\chi }_{\beta }^{D}-\mathbf{\Omega }_{m}\left[
\gamma ^{m}\right] _{BD}\mathbf{\psi }_{\alpha }^{B}\mathbf{\psi }_{\beta
}^{D})P^{\alpha \beta }\left( \mathbf{\chi \psi }\right) ^{2}\mathbf{\phi }%
_{b}+\alpha _{8}\Phi  \nonumber \\
&&+\beta \left( \mathbf{\Theta }_{\alpha }^{M}\left[ \gamma ^{a}\right]
^{A}{}_{M}\mathbf{\omega }_{b}\mathbf{\psi }_{\beta }^{B}-\overline{\mathbf{%
\Theta }}_{\alpha }^{M}\left[ \gamma _{b}\right] ^{A}{}_{M}\mathbf{\omega }
^{a}\mathbf{\chi }_{\beta }^{B}\right)  \nonumber \\
&&\times Q_{AB}P^{\alpha \beta }\left( \mathbf{\chi \psi }\right)
^{3}\varepsilon _{acde}\varepsilon ^{bfgh}\mathbf{\omega }_{fgh}\mathbf{%
\omega }^{cde}  \nonumber \\
&&+\lambda _{1}\left( \mathbf{\Pi }_{\alpha \lambda }^{R}+\mathbf{\Pi }%
_{\alpha \lambda }^{I}\right) \mathbf{\chi }_{\beta }^{A}\mathbf{\psi }%
_{\rho }^{B}P^{\lambda \rho }P^{\alpha \beta }Q_{AB}\left( \mathbf{\chi \psi 
}\right) ^{3}\delta _{b}^{a}\varepsilon _{acde}\varepsilon ^{bfgh}\mathbf{%
\omega }_{fgh}\mathbf{\omega }^{cde}\mathbf{\phi }_{f}  \nonumber \\
&&\left. +\lambda _{2}\left( \mathbf{\Pi }_{\alpha \lambda }^{R}+\mathbf{\Pi 
}_{\alpha \lambda }^{I}\right) \left[ \gamma _{5}\right] _{BD}\mathbf{\psi }%
_{\rho }^{B}\mathbf{\chi }_{\beta }^{D}P^{\rho \lambda }P^{\alpha \beta
}\left( \mathbf{\chi \psi }\right) ^{2}\mathbf{\phi }_{b}\right\}
\label{Action}
\end{eqnarray}
where 
\[
\left( \mathbf{\chi \psi }\right) ^{n}=\left( \mathbf{\chi }_{\alpha
}^{A}Q_{AB}P^{\alpha \beta }\mathbf{\psi }_{\beta }^{B}\right) ^{n} 
\]
for any integer, $n$ and where $\alpha _{1},\alpha _{2}...\alpha _{8},\beta
,\lambda _{1}$ and $\lambda _{2}$ are arbitrary constant coefficients. The
most general, linear, odd parity action is given in Appendix $III$.

By integrating over the fermionic degrees of freedom, $S$ reduces to the
most general linear action found in \cite{WW}, together with a generic
matter term of the form $g(x)\mathbf{\phi }_{b}$. This bosonic action is
known to produce general relativity over a $4$-dim subspace \cite{WW}.
Relations to other superconformal actions are discussed in the final section.

\section{The N=1 Case}

We investigate the $N=1$ case in some detail. Since any fermionic $8$-form
is proportional to the volume form, we may define a tensor $\sigma
^{ABCDEFGH}$ by 
\[
\mathbf{\chi }^{A}\mathbf{\chi }^{B}\mathbf{\chi }^{C}\mathbf{\chi }^{D}%
\mathbf{\psi }^{E}\mathbf{\psi }^{F}\mathbf{\psi }^{G}\mathbf{\psi }%
^{H}=\sigma ^{ABCDEFGH}\phi _{f} 
\]
Therefore, 
\[
\sigma ^{ABCDEFGH}\equiv \left| Q^{A\left[ E\right. }Q^{\left| B\right|
F}Q^{\left| C\right| G}Q^{\left| D\right| \left. H\right] }\right| 
\]
Note that the antisymmetrization removes the ``diagonal'' terms as required
by the discussion of the previous section, while the absolute value restores
symmetry under interchanges. It is also convenient to define 
\[
\sigma _{\qquad B_{1}\ldots B_{k}}^{A_{1}\ldots A_{k}}\equiv \sigma _{\qquad
B_{1}\ldots B_{k}A_{k+1}\ldots A_{N}}^{A_{1}\ldots A_{k}A_{k+1}\ldots A_{N}}=%
\frac{k!\left( n-k\right) !}{n!}\left| \delta _{B_{1}\ldots
B_{k}}^{A_{1}\ldots A_{k}}\right| 
\]

Again, we wish to consider gauge invariant actions over the superspace,
linear in the curvature two forms. The curvatures under consideration are
the seven two-forms, 
\[
\left\{ \mathbf{\Omega }_{b}^{a},\mathbf{\Omega }^{a},\mathbf{\Omega }_{a},%
\mathbf{\Omega },\mathbf{A},\mathbf{\Theta }^{B},\overline{\mathbf{\Theta }}%
^{B}\right\} 
\]
Then replacing $P_{\alpha \beta }\rightarrow i$ in eq.(\ref{Action}), the
most general, gauge-invariant, even parity, action linear in the curvatures
is

\begin{eqnarray}
S &=&\int \left\{ \left( \alpha _{1}\mathbf{\Omega }_{b}^{a}\left[ \sigma
_{a}^{b}\right] _{AB}+(\alpha _{2}\mathbf{\Omega }+\alpha _{3}\mathbf{A}
)Q_{AB}\right) \mathbf{\chi }^{A}\mathbf{\psi }^{B}\left( \mathbf{\chi \psi }%
\right) ^{2}\mathbf{\phi }_{b}\right.  \nonumber \\
&&+(\alpha _{4}\mathbf{\Omega }_{b}^{a}+(\alpha _{5}\mathbf{\Omega }+\alpha
_{6}\mathbf{A})\delta _{b}^{a})\varepsilon _{acde}\varepsilon ^{bfgh}\mathbf{
\omega }_{fgh}\mathbf{\omega }^{cde}\mathbf{\phi }_{f}+\alpha _{7}\Phi 
\nonumber \\
&&+\alpha _{8}(\mathbf{\Omega }^{m}\left[ \gamma _{m}\right] _{BD}\mathbf{%
\chi }^{B}\mathbf{\chi }^{D}-\mathbf{\Omega }_{m}\left[ \gamma ^{m}\right]
_{BD}\mathbf{\psi }^{B}\mathbf{\psi }^{D})\left( \mathbf{\chi \psi }\right)
^{2}\mathbf{\phi }_{b}  \nonumber \\
&&+\beta _{1}\mathbf{\Theta }^{M}\left[ \gamma ^{a}\right] ^{A}{}_{M}\mathbf{%
\omega }_{b}\mathbf{\psi }^{B}Q_{AB}\left( \mathbf{\chi \psi }\right)
^{3}\varepsilon _{acde}\varepsilon ^{bfgh}\mathbf{\omega }_{fgh}\mathbf{\
\omega }^{cde}  \nonumber \\
&&-\beta _{1}\overline{\Theta }^{M}\left[ \gamma _{b}\right] ^{A}{}_{M}%
\mathbf{\omega }^{a}\mathbf{\chi }^{B}\left. Q_{AB}\left( \mathbf{\chi \psi }%
\right) ^{3}\varepsilon _{acde}\varepsilon ^{bfgh}\mathbf{\omega }_{fgh}%
\mathbf{\omega }^{cde}\right\}  \label{Even Parity}
\end{eqnarray}
We next examine some properties of the field equations for $S.$

\section{The Field Equations}

Twenty-eight tensor equations result from variation of the action with
respect to the seven one-forms $\mathbf{\omega }_{a},\mathbf{\omega }^{a},%
\mathbf{\psi }^{B},\mathbf{\chi }^{B},\mathbf{\omega },\mathbf{\omega }%
_{b}^{a}$ and $\mathbf{\alpha }$. From the discussion of the gauge
transformations we recall that under local symmetry transformations the
bosonic and fermionic curvatures do not mix. This observation naturally
gives rise to the question: how does the supersymmetry of the model appear?
The answer lies in the field equations. The field equations generated by the
action relate the components of the bosonic curvatures, $\mathbf{\Omega }%
_{a},\mathbf{\Omega }^{a},\mathbf{\Omega },\mathbf{\Omega }_{b}^{a},\mathbf{%
A,}$ with those of the fermionic curvatures $\mathbf{\Theta }^{A},\overline{%
\mathbf{\Theta }}^{A}$, and under general coordinate transformations the
components of the two types of curvatures will mix.

For example, there are four equations generated by varying the action with
respect to the solder form, $\mathbf{\omega }^{a}.$ One of these is: 
\begin{eqnarray}
0 &=&144\alpha _{1}\Omega _{l}^{m}{}_{n\widetilde{N}}\left[ \sigma
_{m}^{l}\right] _{AB}\sigma ^{AM\left. B\widetilde{N}\right. }-144\alpha
_{2}\Omega {}_{n\widetilde{N}}\sigma ^{M\left. \widetilde{N}\right.
}-144\alpha _{3}A{}_{n\widetilde{N}}\sigma ^{M\left. \widetilde{N}\right. } 
\nonumber \\
&&-2\alpha _{8}\overline{\Theta }^{A}{}_{G\widetilde{H}}\left[ \gamma
_{n}\right] {}_{BD}\sigma ^{MD\left. G\widetilde{H}\right. }-\alpha _{8}%
\overline{\Theta }^{E}{}_{\widetilde{G}\widetilde{H}}\left[ \gamma
_{n}\right] {}_{BD}\sigma ^{BND\left. G\widetilde{H}\right. }{}_{E} 
\nonumber \\
&&-144\alpha _{8}\Omega {}_{mnN}\left[ \gamma ^{m}\right] _{BD}\sigma
^{MN\left. BD\right. }-48\beta _{1}\Theta ^{L}{}_{an}\left[ \gamma
^{a}\right] ^{A}{}_{L}\sigma ^{M}{}_{A}  \label{Rarita Schw}
\end{eqnarray}
\qquad We see that in this expression, one component of the spacetime
curvature tensor, $\Omega _{l}^{m}{}_{n\widetilde{N}},$ is related to the
fermionic curvature tensors. We can therefore eliminate certain fermionic 
\textit{components} of the spacetime curvature in favor of the fermionic 
\textit{curvatures}. Since supercoordinate transformations mix the bosonic
and fermionic parts of the spacetime curvature, the different curvatures
must mix. Similar comments apply to the torsion, co-torsion and dilatation.

Note that eq.(\ref{Rarita Schw}), a fermionic piece of the Einstein
equation, is a Rarita-Schwinger type equation since the final term is
proportional to 
\begin{eqnarray*}
\left[ \gamma ^{a}\right] ^{M}{}_{L}\Theta ^{L}{}_{an} &\sim &\gamma
^{a}\left( \partial _{a}\psi _{n}-\partial _{n}\psi _{a}\right) \\
&\sim &\partial \hspace{-0.1in}/\psi _{n}
\end{eqnarray*}
where we have supressed the spinor index on $\psi $ and $\gamma $. Thus, the
special cases with $\alpha _{1}=\alpha _{2}=\alpha _{8}=0$ give the massless
Dirac equation for a spin-3/2 particle.

\section{Conclusion}

We have formulated the biconformal supergravity theory of the superconformal
group, $SU(2,2|N),$ writing the most general even and odd parity action
linear in the curvatures. The result is a $16$-dim superspace with local
Lorentz and dilatational symmetries. Finding the field equations for the $N=1
$ case illustrates how supercoordinate transformations will mix the
fermionic and bosonic curvatures.

These results are important for several reasons:

\begin{enumerate}
\item  The $N=5$ case is a gauging of the $AdS_{5}\times S^{5}$ background
of the Maldacena conjecture, and therefore provides (at least) the linear
curvature or low energy limit of the string theory of the conjecture.

\item  There is a large class of actions which are linear in the
supercurvatures without auxiliary fields, permitting (1) GR-type gravity
theory and (2) Dirac-type and Rarita-Schwinger-type spinor equations.

\item  Supersymmetrization introduces matter systematically into biconformal
space.

\item  Our use of Cartan's group manifold methods gives superspace
automatically.
\end{enumerate}

Two of these points merit further discussion. We begin by comparing the
class of curvature-linear actions we have written with the actions used by
previous authors. Then we comment briefly on the relationship between the
Maldacena conjecture and $SU(2,2|5)$ biconformal supergravity.

Numerous papers (\cite{CastFre}, \cite{Freund1}-\cite{TvN2}) have examined
properties of conformal supergravity in four dimensions. With the exception
of \cite{CrispimRomao} and \cite{CastFre}, all of these authors use actions
quadratic in the curvatures, typically making use of the MacDowell-Mansouri 
\cite{MacDowell} approach.

For example, the MacDowell-Mansouri approach is used by Crispim-Romao,
Ferber and Freund \cite{CrispimRomao}, who write a curvature squared action 
\begin{eqnarray*}
A_{1} &=&\int d^{4}x\varepsilon ^{\mu \nu \rho \sigma }R_{\mu \nu
}^{A}R_{\rho \sigma }^{B}M_{AB} \\
&=&\int d^{4}x\varepsilon ^{\mu \nu \rho \sigma }R_{\quad \mu \nu
}^{ab}R_{\quad \rho \sigma }^{cd}\varepsilon _{abcd}
\end{eqnarray*}
where the indices $A,B=1,\ldots ,15$ range over all generators of the
conformal group, including special conformal transformations. In this
approach, the curvature is expanded in terms of a reduced symmetry group and
the leading quadratic term becomes topological. The action is then
essentially the Einstein-Hilbert action plus a cosmological term. These
authors also consider the case of only Lorentz and dilatational gauge
fields, with a linear curvature action and linear torsion, coupled to an
auxiliary tensor $\psi ^{AB}$, all in superspace: 
\[
A_{2}=\int d^{4+4N}z\det \left( e_{M}^{\quad A}\right) \left( \left(
-1\right) ^{b+bc}\psi ^{CA}R_{ABC}^{\quad \quad B}+\left( -1\right) ^{a}\mu
D_{A}\psi ^{BC}T_{CB}^{\quad \quad A}\right) 
\]
They modify this, replacing $\psi ^{AB}$ with two copies of a vectorial
superfield $\chi ^{A}$ to eliminate a second derivative term.

Ferrara, Kaku, Townsend and van Nieuwenhuizen (\cite{KTvN1}-\cite{TvN2})
begin with the same MacDowell-Mansouri action, $A_{1},$ including torsion,
co-torsion, dilatation and internal $U(1)$ quadratic terms: 
\begin{eqnarray*}
I &=&\int d^{4}x\varepsilon ^{\mu \nu \rho \sigma }\left( \alpha R_{\quad
\mu \nu }^{ab}R_{\quad \rho \sigma }^{cd}\varepsilon _{abcd}\right. \\
&&+\beta R_{\mu \nu }^{\alpha }\left( Q\right) \left( \gamma _{5}C\right)
_{\alpha \beta }R_{\rho \sigma }^{\beta }\left( S\right) +\left. \gamma
R_{\mu \nu }\left( D\right) R_{\rho \sigma }\left( A\right) \right)
\end{eqnarray*}
In addition, they demand vanishing torsion and self dual gravitino field, 
\begin{eqnarray*}
0 &=&R_{\mu \nu }^{a}\left( P\right) \\
0 &=&R_{\mu \nu }^{n}\left( Q\right) +\frac{1}{2}\gamma _{5}\tilde{R}_{\mu
\nu }^{n}\left( Q\right)
\end{eqnarray*}
When the constraints are substituted, $I$ develops a term linear in the
curvature and also terms built from additional fields. In \cite{KTVN4}, the $
4$-dim action is shown to be invariant under all of the remaining
superconformal transformations.

In all of these papers except \cite{CrispimRomao}, the integrals are four
dimensional, so the local gauge group is the co-Poincar\'{e} group. The
gauge transformations therefore mix essentially all of the curvatures. This
is in sharp constrast to our action, since the group quotient method
requires only local Lorentz and local dilatational invariance.

Only \cite{CrispimRomao} and \cite{CastFre} claim actions which are linear
in the curvatures. This is accomplished using the auxiliary $H$-invariant
tensors that exist by virtue of the reduced local symmetry (from
co-Poincar\'{e} to homothetic) combined with the demand that $R^{A}=0$ solve
the field equations. It seems likely that these fields may be derived from
the biconformal approach by integrating over the extra $4$ coordinates.

Finally, the classes of biconformal action we present in eqs.(\ref{Even
Parity}) and (\ref{Odd parity}) are constructed purely from the geometry
without auxiliary fields. The supersymmetry is carried entirely by
supercoordinate transformations in the underlying $16$-dim superspace.

Next, we comment on conformal supergravity as the low energy limit of string
theory on $AdS_{5}\times S^{5}.$ Since string theory is free of ghosts while
most conformal gauge theories are not, such a relationship might be thought
impossible. However, our linear curvature action removes this obstacle, and
the issue must be examined in further detail. Here we discuss some features
of such a possible correspondence.

Normally, gravitational gauge theories start with the symmetry of a highly
symmetric space. Gauging then leads to a class of geometries closely related
to the first. For example, the gauge theory of the Poincar\'{e} symmetry of Minkowski spacetime leads to the class of pseudo-Riemannian spacetimes. Each of these spacetimes has a copy of the original Minkowski space as the tangent space
at each point. The connection is then clear: the gauge theory is a
perturbation, or deformation, of the original space.

With this in mind, the standard gaugings of the superconformal group are
expected to give spaces that are locally $AdS_{5}\times S^{5}$. Indeed, the
internal $S^{5}$ is not disturbed in these models -- the gauged spacetimes
retain a copy of $S^{5}$ at each point. The $AdS_{5}$ becomes curved in a
way that depends on the field content.

Biconformal gauging is different from the standard gauging, because it
doubles the dimension of the bosonic base space. Therefore, instead of a $10$
-dim generalization of $AdS_{5}\times S^{5},$ the gauging leads to a
superspace with $\left( 8+5\right) $-dimensional bosonic sector. The
question naturally arises, what is this space? Clearly, $5$ dimensions
reflect the $SU(5)$ internal symmetry. To understand the meaning of the
remaining $8$ dimensions, consider the bosonic biconformal spaces studied in 
\cite{NCG} and \cite{WW}. In these cases, biconformal space is found to have
symplectic structure relating coordinates with opposite scaling dimension
(note the similarity to the $U$ coordinate with dimensions of mass in
Maldacena \cite{Maldacena}). The interpretation of biconformal space as a
generalization of phase space has proved quite successful. Indeed, applying
the technique to a conformally invariant generalization of Newton's second
law \cite{GaugeNewton} produces Hamiltonian dynamics as a gauge theory.
Therefore, despite the increased dimension, we still expect the biconformal
supergravity theory to describe curved $AdS_{5}\times S^{5}$ on a
submanifold, with the remaining dimensions providing momentum information.
The exact character of this additional momentum information will be the
subject of further study.

The introduction of string into the $AdS_{5}\times S^{5}$ background poses
another problem. How does string move in the biconformal superspace? The
only previous study of matter in biconformal space \cite{WWMatter} shows
that scalar fields which are $a$ $priori$ dependent on all $2n$-dimensions
of biconformal space reduce under the field equations to fields defined on $
n $-dimensions satisfying the usual $n$-dim scalar field equations. We
conjecture that string is similarly constrained by its equations of motion
and interactions with the curvatures to its usual $AdS_{5}\times S^{5}$
motions. Once again, further study is required before we have a complete
answer\smallskip \bigskip

One of us (JTW) would like to thank P.G.O.Freund for pointing out an error
in an early version of this work, and P. van Nieuwenhuizen, E. Witten and Y.
S. Wu for stimulating discussions.\pagebreak \noindent 

\noindent {\Large Appendix I: Superconformal generators and Lie algebra}

\medskip

Let a generic element of $su(2,2|N)$ be written as 
\[
T=\left( 
\begin{array}{cc}
A & B \\ 
C & D
\end{array}
\right) 
\]
where 
\begin{eqnarray}
QA+A^{\dagger }Q &=&0 \\
C &=&-P^{-1}B^{\dagger }Q \\
PD+D^{\dagger }P &=&0
\end{eqnarray}
Identifying $A$-type conformal generators with $B=C=D=0,$ $G$-type fermionic
generators with $A=D=0,$ and $D$-type generators with $A=B=C=0,$ we may
choose the $A$-type generators as in eqs.(\ref{M}-\ref{D}), the $D$-type as

\begin{eqnarray*}
\left[ D_{R}^{\alpha \beta }\right] _{\quad \nu }^{\mu } &=&i\left( P^{\mu
\alpha }\delta _{\nu }^{\beta }-P^{\mu \beta }\delta _{\nu }^{\alpha
}\right)  \\
\left[ D_{I}^{\alpha \beta }\right] _{\quad \nu }^{\mu } &=&P^{\mu \alpha
}\delta _{\nu }^{\beta }+P^{\mu \beta }\delta _{\nu }^{\alpha }
\end{eqnarray*}
where we have chosen,
\[
P_{\alpha \beta }=\left( 
\begin{array}{cc}
i1 &  \\ 
& i1
\end{array}
\right) 
\]
and the fermionic generators as 
\[
\left[ G_{A}^{\alpha -}\right] =\left( 
\begin{array}{cc}
& \frac{1}{2}\delta _{\beta }^{\alpha }\left[ 1+\gamma _{5}\right] _{\quad
A}^{B} \\ 
\frac{1}{2}P^{\alpha \beta }Q_{BC}\left[ 1-\gamma _{5}\right] _{\quad
A}^{B\,} & 
\end{array}
\right) 
\]
and 
\[
\left[ G_{A}^{\alpha +}\right] =\left( 
\begin{array}{cc}
& \frac{1}{2}\delta _{\beta }^{\alpha }\left[ 1-\gamma _{5}\right] _{\quad
A}^{B} \\ 
\frac{1}{2}P^{\alpha \beta }Q_{BC}\left[ 1+\gamma _{5}\right] _{\quad
A}^{B\,} & 
\end{array}
\right) 
\]
where $\alpha ,\beta =1...N$ and $A,B=1...4.$

The Lie algebra is as follows. For the symplectic $\left( A\right) $ sector
one finds the Lie algebra of the conformal group, 
\begin{eqnarray*}
\left[ M_{\quad b}^{a},M_{\quad d}^{c}\right] &=&\frac{1}{2}\left( \delta
_{b}^{c}\delta _{e}^{a}\delta _{d}^{f}\right. -\eta ^{ac}\eta _{be}\delta
_{d}^{f}-\eta _{bd}\eta ^{cf}\delta _{e}^{a}+\eta _{be}\eta ^{cf}\delta
_{d}^{a} \\
&&-\eta ^{af}\eta _{de}\delta _{b}^{c}+\eta ^{ac}\eta _{de}\delta
_{b}^{f}+\eta _{bd}\eta ^{af}\delta _{e}^{c}-\left. \delta _{b}^{f}\delta
_{e}^{c}\delta _{d}^{a}\right) M_{\quad f}^{e} \\
\left[ M_{\quad b}^{a},P_{c}\right] &=&\eta _{bc}\eta ^{ad}P_{d}-\delta
_{c}^{a}P_{b}=-2\Delta _{cb}^{ad}P_{d} \\
\left[ M_{\quad b}^{a},K^{c}\right] &=&\delta _{b}^{c}K^{a}-\eta _{bd}\eta
^{ac}K^{d}=2\Delta _{db}^{ac}K^{d}
\end{eqnarray*}
\[
\begin{array}{cc}
\left[ P_{a},K^{b}\right] =-2\Delta _{ca}^{bd}M_{\quad d}^{c}-2\delta
_{a}^{b}D & \hspace{0.25in}\left[ D,P_{a}\right] =-P_{a} \\ 
\left[ K^{a},P_{b}\right] =2\Delta _{db}^{ac}M_{\quad c}^{d}+2\delta
_{b}^{a}D & \hspace{0.25in}\left[ D,K^{a}\right] =K^{a}
\end{array}
\]
The unitary $D$-sector has the commutation relations 
\begin{eqnarray*}
\left[ D_{R}^{\alpha \beta },D_{R}^{\mu \nu }\right] &=&\frac{1}{2}\left(
iP^{\beta \mu }\delta _{\rho }^{\alpha }\delta _{\sigma }^{\nu }\right.
-iP^{\beta \nu }\delta _{\rho }^{\alpha }\delta _{\sigma }^{\mu }-iP^{\alpha
\mu }\delta _{\rho }^{\beta }\delta _{\sigma }^{\nu }+iP^{\alpha \nu }\delta
_{\rho }^{\beta }\delta _{\sigma }^{\mu } \\
&&-iP^{\beta \mu }\delta _{\sigma }^{\alpha }\delta _{\rho }^{\nu
}+iP^{\beta \nu }\delta _{\sigma }^{\alpha }\delta _{\rho }^{\mu
}+iP^{\alpha \mu }\delta _{\sigma }^{\beta }\delta _{\rho }^{\nu }-\left.
iP^{\alpha \nu }\delta _{\sigma }^{\beta }\delta _{\rho }^{\mu }\right)
D_{R}^{\rho \sigma } \\
\left[ D_{R}^{\alpha \beta },D_{I}^{\mu \nu }\right] &=&\frac{1}{2}\left(
iP^{\beta \mu }\delta _{\rho }^{\alpha }\delta _{\sigma }^{\nu }\right.
+iP^{\beta \nu }\delta _{\rho }^{\alpha }\delta _{\sigma }^{\mu }-iP^{\alpha
\mu }\delta _{\rho }^{\beta }\delta _{\sigma }^{\nu }-iP^{\alpha \nu }\delta
_{\rho }^{\beta }\delta _{\sigma }^{\mu } \\
&&+iP^{\beta \mu }\delta _{\sigma }^{\alpha }\delta _{\rho }^{\nu
}+iP^{\beta \nu }\delta _{\sigma }^{\alpha }\delta _{\rho }^{\mu
}-iP^{\alpha \mu }\delta _{\sigma }^{\beta }\delta _{\rho }^{\nu }-\left.
iP^{\alpha \nu }\delta _{\sigma }^{\beta }\delta _{\rho }^{\mu }\right)
D_{I}^{\rho \sigma } \\
\left[ D_{I}^{\alpha \beta },D_{R}^{\mu \nu }\right] &=&\frac{1}{2}\left(
iP^{\beta \mu }\delta _{\rho }^{\alpha }\delta _{\sigma }^{\nu }\right.
-iP^{\beta \nu }\delta _{\rho }^{\alpha }\delta _{\sigma }^{\mu }+iP^{\alpha
\mu }\delta _{\rho }^{\beta }\delta _{\sigma }^{\nu }-iP^{\alpha \nu }\delta
_{\rho }^{\beta }\delta _{\sigma }^{\mu } \\
&&+iP^{\beta \mu }\delta _{\sigma }^{\alpha }\delta _{\rho }^{\nu
}-iP^{\beta \nu }\delta _{\sigma }^{\alpha }\delta _{\rho }^{\mu
}+iP^{\alpha \mu }\delta _{\sigma }^{\beta }\delta _{\rho }^{\nu }-\left.
iP^{\alpha \nu }\delta _{\sigma }^{\beta }\delta _{\rho }^{\mu }\right)
D_{I}^{\rho \sigma } \\
\left[ D_{I}^{\alpha \beta },D_{I}^{\mu \nu }\right] &=&\frac{1}{2}\left(
-iP^{\beta \mu }\delta _{\rho }^{\alpha }\delta _{\sigma }^{\nu }\right.
-iP^{\beta \nu }\delta _{\rho }^{\alpha }\delta _{\sigma }^{\mu }-iP^{\alpha
\mu }\delta _{\rho }^{\beta }\delta _{\sigma }^{\nu }-iP^{\alpha \nu }\delta
_{\rho }^{\beta }\delta _{\sigma }^{\mu } \\
&&+iP^{\beta \mu }\delta _{\sigma }^{\alpha }\delta _{\rho }^{\nu
}+iP^{\beta \nu }\delta _{\sigma }^{\alpha }\delta _{\rho }^{\mu
}+iP^{\alpha \mu }\delta _{\sigma }^{\beta }\delta _{\rho }^{\nu }+\left.
iP^{\alpha \nu }\delta _{\sigma }^{\beta }\delta _{\rho }^{\mu }\right)
D_{R}^{\rho \sigma }
\end{eqnarray*}
Finally, the fermionic generators satisfy 
\[
\begin{array}{ccc}
\left[ D,G_{A}^{\alpha +}\right] =\frac{1}{2}G_{A}^{\alpha +} & \left[
P_{a},G_{A}^{\alpha +}\right] =\delta _{\mu }^{\alpha }\left[ \gamma
_{a}\right] _{\quad A}^{C}\left[ G_{C}^{\mu -}\right] & \left[
P_{a},G_{A}^{\alpha -}\right] =0 \\ 
\left[ D,G_{A}^{\alpha -}\right] =-\frac{1}{2}G_{A}^{\alpha -} & \left[
K^{a},G_{A}^{\alpha -}\right] =\delta _{\mu }^{\alpha }\left[ \gamma
^{a}\right] _{\quad A}^{C}\left[ G_{C}^{\mu +}\right] & \left[
K^{a},G_{A}^{\alpha +}\right] =0
\end{array}
\]
\[
\begin{array}{c}
\left[ M_{\quad b}^{a},G_{A}^{\alpha +}\right] =\delta _{\mu }^{\alpha
}\left[ \sigma _{\quad b}^{a}\right] _{\quad A}^{C}\left[ G_{C}^{\mu
+}\right] \\ 
\left[ M_{\quad b}^{a},G_{A}^{\alpha -}\right] =\delta _{\mu }^{\alpha
}\left[ \sigma _{\quad b}^{a}\right] _{\quad A}^{C}\left[ G_{C}^{\mu
-}\right]
\end{array}
\]
with the conformal generators, 
\begin{eqnarray*}
\left[ E,G_{A}^{\alpha +}\right] &=&-\frac{i\left( N-4\right) }{4N}\left[
\gamma _{5}\right] ^{C}{}_{A}\delta _{\lambda }^{\alpha }\left[
G_{C}^{\lambda +}\right] \\
\left[ E,G_{A}^{\alpha -}\right] &=&+\frac{i\left( N-4\right) }{4N}\left[
\gamma _{5}\right] ^{C}{}_{A}\delta _{\lambda }^{\alpha }\left[
G_{C}^{\lambda -}\right]
\end{eqnarray*}
with the $E$, 
\begin{eqnarray*}
\left[ D_{R}^{\mu \nu },G_{A}^{\alpha +}\right] &=&\left[ \gamma _{5}\right]
_{\quad A}^{C}\left( P^{\alpha \mu }\delta _{\lambda }^{\nu }+P^{\alpha \nu
}\delta _{\lambda }^{\mu }\right) \left[ G_{C}^{\lambda +}\right] \\
\left[ D_{R}^{\mu \nu },G_{A}^{\alpha -}\right] &=&-\left[ \gamma
_{5}\right] _{\quad A}^{C}\left( P^{\alpha \mu }\delta _{\lambda }^{\nu
}+P^{\alpha \nu }\delta _{\lambda }^{\mu }\right) \left[ G_{C}^{\lambda
-}\right] \\
\left[ D_{I}^{\mu \nu },G_{A}^{\alpha +}\right] &=&-i\delta _{A}^{C}\left(
P^{\alpha \mu }\delta _{\lambda }^{\nu }-P^{\alpha \nu }\delta _{\lambda
}^{\mu }\right) \left[ G_{C}^{\lambda +}\right] \\
\left[ D_{I}^{\mu \nu },G_{A}^{\alpha -}\right] &=&i\delta _{A}^{C}\left(
P^{\alpha \mu }\delta _{\lambda }^{\nu }-P^{\alpha \nu }\delta _{\lambda
}^{\mu }\right) \left[ G_{C}^{\lambda -}\right]
\end{eqnarray*}
with the $U(N)$ generators, and 
\begin{eqnarray*}
\left\{ G_{A}^{\alpha +},G_{B}^{\beta -}\right\} &=&\left( \frac{1}{2}%
P^{\alpha \beta }\left( -Q_{AB}D+\left[ \sigma _{\quad b}^{a}\right]
_{AB}M_{\quad a}^{b}\right) \right. \\
&&-\frac{i}{2}Q_{AB}D_{R}^{\alpha \beta }+\frac{1}{2}\left[ \gamma
_{5}\right] _{AB}D_{I}^{\alpha \beta }+\left. \left[ \gamma _{5}\right]
^{C}{}_{B}Q_{CA}P^{\alpha \beta }E\right)
\end{eqnarray*}

\[
\begin{array}{cc}
\left\{ G_{A}^{\alpha +},G_{B}^{\beta +}\right\} =P^{\alpha \beta }\left[
\gamma _{a}\right] _{\left( AB\right) }\left[ K^{a}\right] _{\quad D}^{C} & 
\left\{ G_{A}^{\alpha -},G_{B}^{\beta -}\right\} =P^{\alpha \beta }\left[
\gamma ^{a}\right] _{\left( AB\right) }\left[ P_{a}\right] _{\quad D}^{C}
\end{array}
\]
with one another.

\bigskip

\noindent {\Large Appendix II:}

\medskip

In this Appendix we prove that there are three possible choices of
supersymmetric extension of the local symmetry $H$ which are consistent with
the following two properties:

\begin{enumerate}
\item  The $A$-sector of the bosonic part of $H$ must consist of Lorentz
transformations and dilatations.

\item  The $D$-sector of the bosonic part of $H$ must be $U\left( N\right) ,$
thereby retaining the entire $D$-sector as a local internal symmetry.
\end{enumerate}

The proof is as follows.

As noted above, any subalgebra that preserves 
\[
\alpha Q+\beta Q\gamma _{5}
\]
for any fixed $\alpha ,\beta \neq 0,$ is homothetic. We generalize this
condition to an arbitrary element of $su(2,2|N)$ and impose both conditions
1 and 2. Thus, the subset of superconformal generators leaving any matrix $%
\mathcal{M}$ invariant generates a subgroup of $SU(2,2|N).$ We want this
subgroup to be our isotropy, $H.$ Let the invariant matrix $\mathcal{M}$ and
a generic generator $T$ be given by 
\begin{eqnarray*}
\mathcal{M} &=&\left( 
\begin{array}{cc}
\alpha Q+\beta Q\gamma _{5} & R \\ 
S & J
\end{array}
\right)  \\
T &=&\left( 
\begin{array}{cc}
A & B \\ 
C & D
\end{array}
\right) 
\end{eqnarray*}
Then invariance, 
\[
\mathcal{M}T+T^{\ddagger }\mathcal{M}=0
\]
for all $T$ gives 
\begin{eqnarray}
0 &=&\left( \alpha Q+\beta Q\gamma _{5}\right) A+RC+A^{\dagger }\left(
\alpha Q+\beta Q\gamma _{5}\right) -C^{\dagger }S  \label{Constraint 1} \\
0 &=&AS+JC+B^{\dagger }\left( \alpha Q+\beta Q\gamma _{5}\right) +D^{\dagger
}S  \label{Constraint 2} \\
0 &=&\left( \alpha Q+\beta Q\gamma _{5}\right) B+RD+A^{\dagger }R-C^{\dagger
}J  \label{Constraint 3} \\
0 &=&SB+JD+B^{\dagger }R+D^{\dagger }J  \label{Constraint 4}
\end{eqnarray}
Applying condition 2 with $A=B=C=0$ we see that 
\begin{eqnarray*}
0 &=&D^{\dagger }S \\
0 &=&RD \\
0 &=&JD+D^{\dagger }J
\end{eqnarray*}
Since the $D$-sector generators span all $N$-dim matrices, the first two
equations require $R=S=0,$ while the third will constrain $D$ unless $J$ is
proportional to the unitary metric $P.$ With $R=S=0,$ the first constraint,
eq.(\ref{Constraint 1}), is satisfied for homothetic generators $A,$ while (%
\ref{Constraint 4}) is satisfied in agreement with Property 2 if and only if 
$J=\lambda P.$ Therefore, the invariance equations reduce to 
\begin{equation}
0=\left( \alpha Q+\beta Q\gamma _{5}\right) B-\lambda C^{\dagger }P
\label{Simpler Constraint 2}
\end{equation}
and its adjoint. We are therefore left with conditions on $B$ and $C.$
Replacing $C^{\dagger }=QBP^{-1}$ in eq(\ref{Simpler Constraint 2}), we have 
\begin{eqnarray*}
0 &=&\left( \alpha Q+\beta Q\gamma _{5}\right) B-\lambda QB \\
&=&\left( \left( \alpha -\lambda \right) Q+\beta Q\gamma _{5}\right) B
\end{eqnarray*}
Now, since $B$ is comprised of $N$ spinors, we can treat this as a matrix
equation for each $B.$ Nonzero $B$ requires 
\begin{eqnarray*}
0 &=&\det \left( \left( \alpha -\lambda \right) Q+\beta Q\gamma _{5}\right) 
\\
&=&\left( \beta ^{2}-\left( \alpha -\lambda \right) ^{2}\right) ^{2}
\end{eqnarray*}
so there are three possible cases: 
\begin{eqnarray*}
\left( \alpha -\lambda \right)  &\neq &\beta  \\
\left( \alpha -\lambda \right)  &=&\beta \neq 0 \\
\left( \alpha -\lambda \right)  &=&-\beta \neq 0
\end{eqnarray*}
The consequences of each of these conditions are described in the text.

\bigskip

\noindent {\Large Appendix III: Odd parity action}

\medskip

The most general, linear, odd parity action is given by

\begin{eqnarray}
S &=&\int \left\{ \left( \alpha _{1}\mathbf{\Omega }_{b}^{a}\left[ \sigma
_{a}^{b}\right] _{AB}+\left( \alpha _{2}\mathbf{\Omega }+\alpha _{3}\mathbf{A%
}\right) \left[ \gamma _{5}\right] _{AB}\right) \mathbf{\chi }_{\alpha }^{A}%
\mathbf{\psi }_{\beta }^{B}P^{\alpha \beta }\left( \mathbf{\chi \psi }%
\right) ^{2}\mathbf{\phi }_{b}\right.  \nonumber \\
&&+\left( \alpha _{4}\mathbf{\Omega }_{b}^{a}+\left( \alpha _{5}\mathbf{\
\Omega }+\alpha _{6}\mathbf{A}\right) \delta _{b}^{a}\right) \varepsilon
_{acde}\varepsilon ^{bfgh}\mathbf{\omega }_{fgh}\mathbf{\omega }^{cde}%
\mathbf{\phi }_{f}  \nonumber \\
&&+\alpha _{7}(\mathbf{\Omega }^{m}\left[ \gamma _{5}\gamma _{m}\right] _{BD}%
\mathbf{\ \chi }_{\alpha }^{B}\mathbf{\chi }_{\beta }^{D}-\mathbf{\Omega }%
_{m}\left[ \gamma _{5}\gamma ^{m}\right] _{BD}\mathbf{\psi }_{\alpha }^{B}%
\mathbf{\psi }_{\beta }^{D})P^{\alpha \beta }\left( \mathbf{\chi \psi }%
\right) ^{2}\mathbf{\phi }_{b}+\alpha _{8}\Phi  \nonumber \\
&&+\beta \left( \mathbf{\Theta }_{\alpha }^{M}\left[ \gamma _{5}\gamma
^{a}\right] ^{A}{}_{M}\mathbf{\omega }_{b}\mathbf{\psi }_{\beta }^{B}-%
\overline{\mathbf{\Theta }}_{\alpha }^{M}\left[ \gamma _{5}\gamma
_{b}\right] ^{A}{}_{M}\mathbf{\omega }^{a}\mathbf{\chi }_{\beta }^{B}\right)
\nonumber \\
&&\times Q_{AB}P^{\alpha \beta }\left( \mathbf{\chi \psi }\right)
^{3}\varepsilon _{acde}\varepsilon ^{bfgh}\mathbf{\omega }_{fgh}\mathbf{%
\omega }^{cde}  \nonumber \\
&&+\lambda _{1}\left( \mathbf{\Pi }_{\alpha \lambda }^{R}+\mathbf{\Pi }
_{\alpha \lambda }^{I}\right) \mathbf{\chi }_{\beta }^{A}\mathbf{\psi }
_{\rho }^{B}P^{\lambda \rho }P^{\alpha \beta }Q_{AB}\left( \mathbf{\chi \psi 
}\right) ^{3}\delta _{b}^{a}\varepsilon _{acde}\varepsilon ^{bfgh}\mathbf{\
\omega }_{fgh}\mathbf{\omega }^{cde}\mathbf{\phi }_{f}  \nonumber \\
&&\left. +\lambda _{2}\left( \mathbf{\Pi }_{\alpha \lambda }^{R}+\mathbf{\Pi 
}_{\alpha \lambda }^{I}\right) \left[ \gamma _{5}\right] _{BD}\mathbf{\psi }%
_{\rho }^{B}\mathbf{\chi }_{\beta }^{D}P^{\rho \lambda }P^{\alpha \beta
}\left( \mathbf{\chi \psi }\right) ^{2}\mathbf{\phi }_{b}\right\}
\label{Odd parity}
\end{eqnarray}
As with the even parity case, there are $11$ arbitrary parameters.

\bigskip

\noindent {\Large Appendix IV: Field equations}

\medskip

The full set of field equations from the $N=1,$ even parity action is as
follows.

From the $\omega $ variation we have, 
\begin{eqnarray*}
0 &=&-3\alpha _{2}\overline{\Theta }^{A}{}_{M\widetilde{H}}\sigma ^{NM\left. 
\widetilde{H}\right. }{}_{A}-\frac{3}{2}\alpha _{2}\Theta ^{B}{}_{\widetilde{%
M}\widetilde{H}}Q_{AB}\sigma ^{AN\left. \widetilde{M}\widetilde{H}\right. }
\\
&&+576\alpha _{2}\Omega _{b}{}^{b}{}_{\widetilde{M}}\sigma ^{N\left. 
\widetilde{M}\right. }{}+576\alpha _{2}\Omega {}^{a}{}_{a\widetilde{M}%
}\sigma ^{N\left. \widetilde{M}\right. } \\
&&+144\alpha _{5}\overline{\Theta }^{Am}{}_{m}{}\sigma ^{N}{}_{A}
\end{eqnarray*}
and 
\begin{eqnarray*}
0 &=&576\alpha _{2}\Omega _{nN\widetilde{M}}\sigma ^{N\left. \widetilde{M}%
\right. }+72\alpha _{5}\Omega _{f}{}^{m}{}_{l}\delta _{nm}^{lf}-36\alpha
_{5}\Omega {}^{c}{}_{nc} \\
&&-144\alpha _{5}\overline{\Theta }^{A}{}_{nN}\sigma ^{N}{}_{A}-144\alpha
_{5}\Theta ^{B}{}_{n\widetilde{M}}\sigma ^{A\left. \widetilde{M}\right.
}Q_{AB}
\end{eqnarray*}
together with the complex conjugates of these expressions.

The $\omega _{b}^{a}$ variation gives 
\begin{eqnarray*}
0 &=&\alpha _{1}\left[ \sigma _{m}^{n}\right] _{AB}\left( -\overline{\Theta }%
^{A}{}_{M\widetilde{H}}\sigma ^{NM\left. B\widetilde{H}\right. }-\frac{1}{2}%
\Theta ^{B}{}_{\widetilde{M}\widetilde{H}}\sigma ^{AN\left. \widetilde{M}%
\widetilde{H}\right. }\right. \\
&&-2\overline{\Theta }^{C}{}_{M\widetilde{H}}\sigma ^{ANM\left. B\widetilde{H%
}\right. }{}_{C}-\Theta ^{D}{}_{\widetilde{M}\widetilde{H}}\sigma
^{ACN\left. B\widetilde{M}\widetilde{H}\right. }Q_{CD} \\
&&+\left. 576\Omega _{b}{}^{b}{}_{\widetilde{N}}\sigma ^{AN\left. B%
\widetilde{N}\right. }{}+576\Omega ^{a}{}_{a\widetilde{N}}\sigma ^{AN\left. B%
\widetilde{N}\right. }\right) \\
&&-144\alpha _{4}\left[ \sigma _{m}^{n}\right] _{AB}\overline{\Theta }%
^{Cn}{}_{m}\sigma ^{N}{}_{C}
\end{eqnarray*}
and 
\begin{eqnarray*}
0 &=&576\alpha _{1}\left[ \sigma _{m}^{n}\right] _{AB}\Omega _{pM\widetilde{H%
}}\sigma ^{AM\left. B\widetilde{H}\right. }+72\alpha _{4}\Omega
_{f}{}^{q}{}_{m}\delta _{pq}^{nf} \\
&&-36\alpha _{4}\Omega ^{c}{}_{mc}\delta _{p}^{n}-144\alpha _{4}\overline{%
\Theta }^{C}{}_{mN}\sigma ^{N}{}_{C}\delta _{p}^{n} \\
&&-144\alpha _{4}\Theta ^{D}{}_{m\widetilde{N}}\sigma ^{C\left. \widetilde{N}%
\right. }Q_{CD}\delta _{p}^{n}
\end{eqnarray*}
and the complex conjugates of these expressions.

For the $\alpha $ variation, the field equations are 
\begin{eqnarray*}
0 &=&-\frac{3}{2}\alpha _{3}\overline{\Theta }^{A}{}_{GH}\sigma ^{GH\left. 
\widetilde{E}\right. }{}_{A}-3\alpha _{3}\Theta ^{B}{}_{G\widetilde{H}%
}\sigma ^{GA\left. \widetilde{E}\widetilde{H}\right. }{}Q_{AB} \\
&&+576\alpha _{3}\Omega _{b}{}^{b}{}_{G}\sigma ^{G\left. \widetilde{E}%
\right. }+576\alpha _{3}\Omega {}_{\quad m}^{m}{}_{G}\sigma ^{G\left. 
\widetilde{E}\right. } \\
&&+144\alpha _{6}\Theta ^{Bm}{}_{m}\sigma ^{A\left. \widetilde{E}\right.
}Q_{AB}
\end{eqnarray*}
and 
\begin{eqnarray*}
0 &=&576\alpha _{3}\Omega _{mG\widetilde{H}}\sigma ^{G\left. \widetilde{H}%
\right. }{}+72\alpha _{6}\Omega _{f}{}^{n}{}_{b}\delta _{mn}^{bf}{} \\
&&-36\alpha _{6}\Omega ^{c}{}_{mc}-144\alpha _{6}\overline{\Theta }%
^{A}{}_{mE}\sigma ^{E\left. B\right. }Q_{AB} \\
&&+144\alpha _{6}\Theta ^{B}{}_{m\widetilde{E}}\sigma ^{A\left. \widetilde{E}%
\right. }Q_{AB}
\end{eqnarray*}
together with conjugate equations.

Variation of $\omega ^{a}$ leads to 
\begin{eqnarray*}
0 &=&144\alpha _{1}\Omega _{l}^{m}{}_{M\widetilde{N}}\left[ \sigma
_{m}^{l}\right] _{AB}\sigma ^{AM\left. B\widetilde{N}\right. }\delta
_{n}^{p}+144\alpha _{2}\Omega {}_{M\widetilde{N}}\sigma ^{M\left. \widetilde{%
N}\right. }\delta _{n}^{p} \\
&&+144\alpha _{3}A{}_{M\widetilde{N}}\sigma ^{M\left. \widetilde{N}\right.
}\delta _{n}^{p}-108\alpha _{4}\delta _{n}^{p}+72\alpha _{5}\delta
_{n}^{p}+24\alpha _{4}\Omega _{b}^{a}{}^{b}{}_{l}\delta _{an}^{pl} \\
&&+24\alpha _{5}\Omega {}^{a}{}_{l}\delta _{an}^{pl}+24\alpha
_{6}A{}^{a}{}_{l}\delta _{an}^{pl}+144\alpha _{7}\delta _{n}^{p} \\
&&+288\alpha _{8}\Omega ^{p}{}{}_{\widetilde{M}\widetilde{H}}\left[ \gamma
_{n}\right] _{BD}\sigma ^{BD\left. \widetilde{M}\widetilde{H}\right.
}+72\alpha _{8}\Omega ^{m}{}{}_{\widetilde{M}\widetilde{N}}\left[ \gamma
_{m}\right] _{BD}\sigma ^{BD\left. \widetilde{M}\widetilde{N}\right. }\delta
_{n}^{p} \\
&&-72\alpha _{8}\Omega {}_{mMN}\left[ \gamma ^{m}\right] _{BD}\sigma
^{MN\left. BD\right. }\delta _{n}^{p}-144\beta _{1}\left[ \gamma ^{p}\right]
^{A}{}_{M}\left[ \gamma _{n}\right] ^{M}{}_{N}\sigma ^{N}{}_{A} \\
&&+96\beta _{1}\Theta ^{M}{}_{mN}\left[ \gamma ^{a}\right] ^{A}{}_{M}\delta
_{an}^{pm}\sigma ^{N}{}_{A}-36\beta _{1}\overline{\Theta }^{Mm}{}_{%
\widetilde{N}}\left[ \gamma _{m}\right] ^{A}{}_{M}\sigma ^{B\left. 
\widetilde{N}\right. }Q_{AB}\delta _{n}^{p}
\end{eqnarray*}
and 
\begin{eqnarray*}
0 &=&144\alpha _{1}\Omega _{l}^{m}{}_{n\widetilde{N}}\left[ \sigma
_{m}^{l}\right] _{AB}\sigma ^{AM\left. B\widetilde{N}\right. }-144\alpha
_{2}\Omega {}_{n\widetilde{N}}\sigma ^{M\left. \widetilde{N}\right.
}-144\alpha _{3}A{}_{n\widetilde{N}}\sigma ^{M\left. \widetilde{N}\right. }
\\
&&-2\alpha _{8}\overline{\Theta }^{A}{}_{G\widetilde{H}}\left[ \gamma
_{n}\right] {}_{BD}\sigma ^{MD\left. G\widetilde{H}\right. }-\alpha _{8}%
\overline{\Theta }^{E}{}_{\widetilde{G}\widetilde{H}}\left[ \gamma
_{n}\right] {}_{BD}\sigma ^{BND\left. G\widetilde{H}\right. }{}_{E} \\
&&-144\alpha _{8}\Omega {}_{mnN}\left[ \gamma ^{m}\right] _{BD}\sigma
^{MN\left. BD\right. }-48\beta _{1}\Theta ^{L}{}_{an}\left[ \gamma
^{a}\right] ^{A}{}_{L}\sigma ^{M}{}_{A}
\end{eqnarray*}
and 
\begin{eqnarray*}
0 &=&144\alpha _{1}\Omega _{l}^{m}{}_{nN}\left[ \sigma _{m}^{l}\right]
_{AB}\sigma ^{AN\left. B\widetilde{M}\right. }-144\alpha _{2}\Omega
{}_{nN}\sigma ^{N\left. \widetilde{M}\right. }\delta _{n}^{p}-144\alpha
_{3}A{}_{nN}\sigma ^{N\left. \widetilde{M}\right. } \\
&&-2\alpha _{8}\overline{\Theta }^{B}{}_{L\widetilde{H}}\left[ \gamma
_{n}\right] {}_{BD}\sigma ^{LD\left. \widetilde{M}\widetilde{H}\right.
}-2\alpha _{8}\overline{\Theta }^{E}{}_{L\widetilde{H}}\left[ \gamma
_{n}\right] {}_{BD}\sigma ^{BDL\left. \widetilde{M}\widetilde{H}\right.
}{}_{E} \\
&&-\alpha _{8}Q_{EF}\Theta ^{F}{}_{\widetilde{L}\widetilde{H}}\left[ \gamma
_{n}\right] {}_{BD}\sigma ^{BDE\left. \widetilde{H}\widetilde{L}\right.
}{}_{A}+576\alpha _{8}\Omega _{b}{}^{b}{}_{\widetilde{L}}\left[ \gamma
_{n}\right] {}_{BD}\sigma ^{BD\left. \widetilde{M}\widetilde{L}\right. } \\
&&+576\alpha _{8}\Omega ^{a}{}_{a\widetilde{H}}\left[ \gamma _{n}\right]
{}_{BD}\sigma ^{BD\left. \widetilde{M}\widetilde{H}\right. }+144\alpha
_{8}\Omega ^{m}{}_{n\widetilde{N}}\left[ \gamma _{m}\right] {}_{BD}\sigma
^{BD\left. \widetilde{M}\widetilde{H}\right. } \\
&&+36\beta _{1}\overline{\Theta }^{Lb}{}_{n}\left[ \gamma _{b}\right]
{}_{BL}\sigma ^{B\left. \widetilde{M}\right. }
\end{eqnarray*}
and 
\begin{eqnarray*}
0 &=&-12\alpha _{4}\Omega _{p}^{a}{}_{an}-12\alpha _{5}\Omega
{}_{pn}-12\alpha _{6}A{}_{pn}+288\alpha _{8}\Omega {}_{p\widetilde{M}%
\widetilde{N}}\left[ \gamma _{n}\right] {}_{BD}\sigma ^{BD\left. \widetilde{M%
}\widetilde{N}\right. } \\
&&+36\beta _{1}\overline{\Theta }^{M}{}_{n\widetilde{N}}\left[ \gamma
_{p}\right] ^{A}{}_{M}Q_{AB}\sigma ^{B\left. \widetilde{N}\right. }
\end{eqnarray*}
The complex conjugates of these equations are obtained through the $\omega
_{a}$ variation.

Finally, variation of $\psi ^{A}$ gives, 
\begin{eqnarray*}
0 &=&\frac{3}{2}\alpha _{1}\Omega _{b}^{a}{}_{\widetilde{M}\widetilde{N}%
}\left[ \sigma _{a}^{b}\right] _{AN}\sigma ^{AL\left. \widetilde{M}%
\widetilde{N}\right. }+\frac{3}{2}\alpha _{2}\Omega {}_{\widetilde{M}%
\widetilde{N}}Q_{AN}\sigma ^{AL\left. \widetilde{M}\widetilde{N}\right. } \\
&&+\frac{3}{2}\alpha _{3}Q_{AN}A{}_{\widetilde{M}\widetilde{N}}\sigma
^{AL\left. \widetilde{M}\widetilde{N}\right. }-2\alpha _{8}\Omega {}_{mP%
\widetilde{H}}\left[ \gamma ^{m}\right] {}_{ND}\sigma ^{PL\left. D\widetilde{%
H}\right. } \\
&&-2\alpha _{8}\Omega {}_{mP\widetilde{H}}\left[ \gamma ^{m}\right]
{}_{BD}\sigma ^{PLEG\left. BD\widetilde{H}\right. }Q_{EN}+144\beta
_{1}\left[ \gamma ^{m}\right] ^{A}{}_{N}\Theta ^{B}{}_{m\widetilde{H}}\sigma
^{L\left. \widetilde{H}\right. }Q_{AB} \\
&&+432\beta _{1}\left[ \gamma ^{m}\right] ^{A}{}_{N}\Theta ^{D}{}_{m%
\widetilde{H}}Q_{AB}Q_{CD}\sigma ^{CL\left. B\widetilde{H}\right. }+144\beta
_{1}\left[ \gamma ^{n}\right] _{BN}\Omega _{b}{}^{b}{}_{n}\sigma ^{L\left.
B\right. } \\
&&+144\beta _{1}\left[ \gamma ^{a}\right] ^{A}{}_{N}\Omega ^{c}{}_{ac}\sigma
^{L}{}_{A}+144\beta _{1}\left[ \gamma ^{m}\right] {}_{NM}\Theta ^{M}{}_{m%
\widetilde{H}}\sigma ^{L\left. \widetilde{H}\right. } \\
&&-432\beta _{1}\left[ \gamma ^{m}\right] ^{A}{}_{M}\Theta ^{M}{}_{m%
\widetilde{H}}\sigma ^{LE\left. \widetilde{H}\right. }Q_{EN}-432\beta
_{1}\left[ \gamma ^{m}\right] {}_{BN}\overline{\Theta }_{\quad mH}^{C}\sigma
^{LHBD}Q_{CD}
\end{eqnarray*}
and 
\begin{eqnarray*}
0 &=&-144\alpha _{4}\Omega _{m}^{n}{}^{m}{}_{\widetilde{N}}\sigma ^{A\left. 
\widetilde{N}\right. }Q_{AN}-144\alpha _{5}\Omega {}^{n}{}_{\widetilde{N}
}\sigma ^{A\left. \widetilde{N}\right. }Q_{AN} \\
&&-144\alpha _{6}A{}^{n}{}_{\widetilde{N}}\sigma ^{A\left. \widetilde{N}%
\right. }Q_{AN}-144\beta _{1}\left[ \gamma ^{n}\right] _{BN}\Theta ^{B}{}_{L%
\widetilde{H}}\sigma ^{L\left. \widetilde{H}\right. } \\
&&-432\beta _{1}\left[ \gamma ^{n}\right] _{BN}\Theta ^{D}{}_{L\widetilde{H}
}Q_{CD}\sigma ^{CL\left. B\widetilde{H}\right. }+144\beta _{1}\Omega
_{b}{}^{b}{}_{L}\left[ \gamma ^{n}\right] ^{A}{}_{N}\sigma ^{L}{}_{A} \\
&&+288\beta _{1}\left[ \gamma ^{a}\right] ^{A}{}_{N}\Omega ^{c}{}_{lL}\sigma
^{L}{}_{A}\delta _{ac}^{nl}+144\beta _{1}\left[ \gamma ^{n}\right]
{}_{NM}\Theta ^{M}{}_{L\widetilde{H}}\sigma ^{L\left. \widetilde{H}\right. }
\\
&&+576\beta _{1}\left[ \gamma ^{n}\right] ^{A}{}_{M}\Theta ^{M}{}_{L%
\widetilde{H}}\sigma ^{LE}{}_{A}Q_{EN}+432\beta _{1}\left[ \gamma
^{n}\right] {}_{BN}\overline{\Theta }_{\quad LH}^{C}\sigma ^{LHB}{}_{C}
\end{eqnarray*}
and 
\begin{eqnarray*}
0 &=&144\alpha _{4}\Omega _{n}^{m}{}_{m\widetilde{N}}\sigma ^{A\left. 
\widetilde{N}\right. }Q_{AN}+144\alpha _{5}\Omega {}_{n\widetilde{N}}\sigma
^{A\left. \widetilde{N}\right. }Q_{AN}-144\alpha A{}_{n\widetilde{N}}\sigma
^{A\left. \widetilde{N}\right. }Q_{AN} \\
&&-144\beta _{1}\left[ \gamma ^{a}\right] ^{A}{}_{N}\Omega _{naL}\sigma
^{L}{}_{A}-432\beta _{1}\left[ \gamma _{n}\right] {}_{BM}\overline{\Theta }%
^{M}{}_{\widetilde{L}\widetilde{N}}\sigma ^{BE\left. \widetilde{L}\widetilde{%
N}\right. }Q_{EN}
\end{eqnarray*}
and 
\begin{eqnarray*}
0 &=&\left( 2i\alpha _{1}\left[ \sigma _{m}^{n}\right] _{AB}\left[ \sigma
_{n}^{m}\right] _{MN}-\frac{i}{2}\alpha _{2}Q_{MN}Q_{AB}+\left[ \gamma
_{5}\right] _{MN}Q_{AB}\right) \sigma ^{AM\left. B\widetilde{N}\right. } \\
&&+3\alpha _{1}\left[ \sigma _{b}^{a}\right] _{AN}\Omega _{aL\widetilde{M}%
}^{b}{}\sigma ^{AL\left. \widetilde{M}\widetilde{N}\right. }+3\alpha
_{2}\Omega {}_{L\widetilde{M}}Q_{AN}\sigma ^{AL\left. \widetilde{M}%
\widetilde{N}\right. } \\
&&+3\alpha _{3}A{}_{L\widetilde{M}}Q_{AN}\sigma ^{AL\left. \widetilde{M}%
\widetilde{N}\right. }-144\alpha _{4}\Omega _{m\quad n}^{n\;m}{}\sigma
^{A\left. \widetilde{N}\right. }Q_{AN} \\
&&+144\alpha _{5}\Omega ^{n}{}{}_{n}\sigma ^{A\left. \widetilde{N}\right.
}Q_{AN}-144\alpha _{6}A^{n}{}{}_{n}\sigma ^{A\left. \widetilde{N}\right.
}Q_{AN} \\
&&+4\alpha _{7}Q_{AN}\sigma ^{A\left. \widetilde{N}\right. }+\alpha _{8}%
\frac{i}{2}\left( \left[ K^{m}\right] _{L\widetilde{N}}+\left[ K^{m}\right]
_{\widetilde{N}L}\right) \left[ \gamma _{m}\right] {}_{BD}\sigma ^{BD\left. 
\widetilde{L}\widetilde{N}\right. } \\
&&+\alpha _{8}\Omega _{\quad \widetilde{L}\widetilde{H}}^{m}{}\left[ \gamma
_{m}\right] {}_{BD}Q_{EN}\sigma ^{BDE\left. \widetilde{L}\widetilde{H}%
\widetilde{N}\right. }-\alpha _{8}\Omega _{mLM}\left[ \gamma ^{m}\right]
{}_{ND}\sigma ^{LM\left. D\widetilde{N}\right. } \\
&&-\alpha _{8}\Omega _{mLM}\left[ \gamma ^{m}\right] {}_{BD}\sigma
^{LME\left. BD\widetilde{N}\right. }Q_{EN}+144\beta _{1}\left[ \gamma
^{m}\right] ^{A}{}_{N}\Theta ^{B}{}_{mL}Q_{AB}\sigma ^{L\left. \widetilde{N}%
\right. } \\
&&+432\beta _{1}\left[ \gamma ^{m}\right] ^{A}{}_{N}\Theta
^{D}{}_{mL}Q_{AB}Q_{CD}\sigma ^{CL\left. B\widetilde{N}\right. } \\
&&-144\beta _{1}\left[ \gamma ^{m}\right] ^{A}{}_{M}\Theta
^{M}{}_{mL}Q_{AN}\sigma ^{L\left. \widetilde{N}\right. }-432\beta _{1}\left[
\gamma ^{m}\right] ^{A}{}_{M}\Theta ^{M}{}_{mL}Q_{EN}\sigma ^{LE\left. 
\widetilde{N}\right. }{}_{A} \\
&&-432\beta _{1}\left[ \gamma _{b}\right] ^{A}{}_{M}\overline{\Theta }%
_{\quad \;\widetilde{L}}^{Mb}{}\sigma ^{BE\left. \widetilde{N}\widetilde{L}%
\right. }Q_{EN}
\end{eqnarray*}
The complex conjugates of these equations are obtained through the $\chi
^{A} $ variation.

\end{document}